\documentclass[final,5p,times,twocolumn]{elsarticle}
\usepackage{color}
\usepackage{booktabs}
\usepackage[table,xcdraw]{xcolor}
\usepackage{xcolor}
\usepackage{multirow}
\usepackage{blindtext}
\usepackage{array}
\usepackage{amsmath}
\usepackage{amssymb}
\usepackage{booktabs}
\usepackage[caption=false]{subfig}
\usepackage{epstopdf}
\usepackage[numbers]{natbib}
\usepackage{lineno,hyperref}
\usepackage{amsthm}
\theoremstyle{plain}
\newtheorem{theorem}{Property}
\newtheorem{defn}{Definition}
\usepackage{ulem}
\usepackage{soul}

\DeclareMathAlphabet      {\mathbfit}{OML}{cmm}{b}{it}
\modulolinenumbers[5]

\journal{Journal of \LaTeX\ Templates}

\renewcommand{\SS}{\mathcal S}
\colorlet{mygreen}{green!60!gray}

\newcommand{\bm}[1]{\mbox{\boldmath{$#1$}}}

\begin{document}
\begin{frontmatter}

\title{CNN-based Steganalysis and Parametric Adversarial Embedding: a Game-Theoretic Framework\tnoteref{mytitlenote}}

\author[CNmainaddress]{Xiaoyu Shi}
\author[ITmainaddress]{Benedetta Tondi}
\author[CNmainaddress,CNsecondaryaddress]{Bin Li\corref{correspondingauthor}}
\cortext[correspondingauthor]{Corresponding author}\ead{libin@szu.edu.cn}
\author[ITmainaddress]{Mauro Barni}


\address[CNmainaddress]{Guangdong Key Laboratory of Intelligent Information Processing and Shenzhen Key Laboratory of Media Security, Shenzhen University, Shenzhen 518060, China}
\address[ITmainaddress]{Department of Information Engineering and Mathematics, University of Siena, Siena 53100, Italy}
\address[CNsecondaryaddress]{Peng Cheng Laboratory, Shenzhen 518052, China}

\begin{abstract}
CNN-based steganalysis has recently achieved very good performance in detecting content-adaptive steganography.
{At the same time, recent works have shown that}, by adopting an approach similar to that used to build adversarial examples, a steganographer can {adopt an adversarial embedding strategy to effectively counter} a target CNN steganalyzer. {In turn}, the good performance of the steganalyzer can be restored by retraining the CNN with adversarial stego images.
{A problem with this model is that, arguably, at training time the steganalizer is not aware of the exact parameters used by the steganograher for adversarial embedding}
{and, vice versa, the steganographer does not know how the images that will be used to train the steganalyzer are generated.}
{In order to exit this apparent deadlock, we introduce a game theoretic framework wherein the problem of setting the parameters of the steganalyzer and the steganographer is solved in a strategic way.}
More specifically, {a non-zero sum game} is first formulated to model the problem, and then instantiated by considering {a specific} adversarial embedding scheme setting its operating parameters in a game-theoretic fashion.
{Our analysis shows that {the equilibrium solution of the non zero-sum game can be conveniently found by solving} an associated zero-sum game, thus reducing greatly the complexity of the problem.}
Then we run several experiments to derive the optimum strategies for
the steganographer and the staganalyst  in a game-theoretic sense, and to
evaluate the performance of the game at the equilibrium, characterizing the loss with respect to the conventional non-adversarial case.
%
Eventually, by leveraging on the analysis of the equilibrium point of the game, we introduce a new strategy to improve the reliability of the steganalysis, which shows the benefits of addressing the security issue in a game-theoretic perspective.
\end{abstract}

\end{frontmatter}

\section{Introduction}
\label{intro}

As a popular technique in multimedia security,
image steganography \cite{fridrich2009steganography,Li2011A} tries to conceal {a secret message in a cover image by slightly modifying} pixel values or DCT coefficients. Accompanying the development of image steganography, steganalysis aims {at detecting the presence of hidden information within an image}. The two techniques {are applied by two competing players} in a hunting and escaping game.

Nowadays, mainstream steganographic  schemes are designed to be content-adaptive under the framework of distortion minimization \cite{Fridrich2007}. {Different schemes are designed by properly choosing the distortion function.} For example, HUGO (Highly Undetectable steGO) \cite{Pevny2010} defines the distortion function according to the impact {that data embedding has on} SPAM (Subtractive Pixel Adjacency Matrix) \cite{Pevny2010a} features. WOW (Wavelet Obtained Weights) \cite{Holub2012} assigns distortion costs by using three wavelet directional filters. S-UNIWARD (Spatial Universal Wavelet Relative Distortion) \cite{Holub2014} is a slightly modified version of WOW, {which can be easily extended to work with JPEG images}. HILL (High-pass, Low-pass, and Low-pass) \cite{Li2014} employs a high-pass and two low-pass filters {to make sure} that pixels within textured regions have relatively low costs.
{Other} schemes {are} designed by minimizing {the difference between cover and stego images characterized by a statistical model}, such as MG (Multivariate Gaussian) \cite{Fridrich2013}, MVGG (Multivariate Generalized Gaussian) \cite{sedighi2015content}, MiPOD (Minimizing the Power of Optimal detector) \cite{Sedighi2016}, and MRG (Multivariate Gaussian
for Residuals) \cite{Qin2019}.
Some useful methods such as MDS (Modification Direction Synchronization) \cite{Li2015,denemark2015improving} and CPP (Controversial Pixels Prior) \cite{Zhou2017a} can be used
to further exploit non-additive distortion.

Steganalysis has also made substantial progress {in this competition}. {The most common approach to counter} content-adaptive steganographic schemes {consists in the analysis of high dimensional feature vectors}  \cite{Fridrich2012,Tang2014,Denemark2014,Holub2013,Li2018}. {The use of} {SRM (Spatial Rich Model)} {features}  \cite{Fridrich2012}  exploiting high-order pixel dependency and resulting in {a feature space with tens of thousands dimensions, is the most representative example of this approach}. {Other} methods {exploit the so called} ``selection-channel'' information. {For instance} ASRM (Adaptive Spatial Rich Models)\cite{Tang2014} and maxSRM \cite{Denemark2014}, put more emphasis {on} the regions {that are more likely modified by the steganographer}.

Motivated by the recent {success of deep-learning in image processing and computer vision applications},  deep-learning-based steganalytic methods relying on Convolution Neural Networks (CNNs) have been explored. Tan-net \cite{Tan2014} was the first deep learning method for steganalysis based on auto-encoders. Later, Qian et al. \cite{Qian2015} proposed a CNN equipped with Gaussian activation functions and high-pass pre-processing filters.
A breakthrough {in this direction has recently been} achieved by Xu et al.\cite{Xu2016}. Their proposed CNN, called Xu-net, is a new structure {explicitly designed for} image steganalysis, which considered several advanced CNN methods for image classification tasks, such as batch normalization (BN) \cite{ioffe2015batch}, Tanh activation function, and $1\times1$ convolution in deeper layer to enhance the strength of modeling. Ye-net \cite{Ye2017} {further improved the performance of CNN-based steganalysis} by incorporating selection-channel information and applying truncated linear unit (TLU) in the first few layers to accelerate {network-training} convergence.
Recently,  an end-to-end deep residual architecture, called SR-net \cite{Boroumand2018}, has been proposed, which minimizes the use of heuristics and hand-crafted components and works well for steganalysis in both spatial and JPEG domains.

\subsection{Motivation}

CNN-based steganalysis can provide better performance with respect to
 standard machine-learning (ML) methods.
However, as shown by many recent works in the general literature of deep learning,  CNNs are vulnerable to so called adversarial examples \cite{szegedy2013intriguing}: slight, often imperceptible, perturbations of
the input, which are sufficient to induce a wrong decision.
The concept of adversarial examples has been already successfully exploited in steganography, to counter a target CNN steganalyzer \cite{zhang2018adversarial,Li2018a,Tang2019}.
In  \cite{Tang2019}, in particular, an adversarial steganographic scheme is developed by adjusting the embedding costs based on the back-propagated gradient {of the target CNN steganalyzer}.\footnote{{Throughout the paper we refer to the person aimed at detecting stego signals as steganalyst. The term steganalyzer is used when we refer to the classifier/network implementing the staganalysis.}}

\begin{figure*}[!t]
\vspace{0.3cm}
	\centering
	\includegraphics[width=0.99\linewidth]{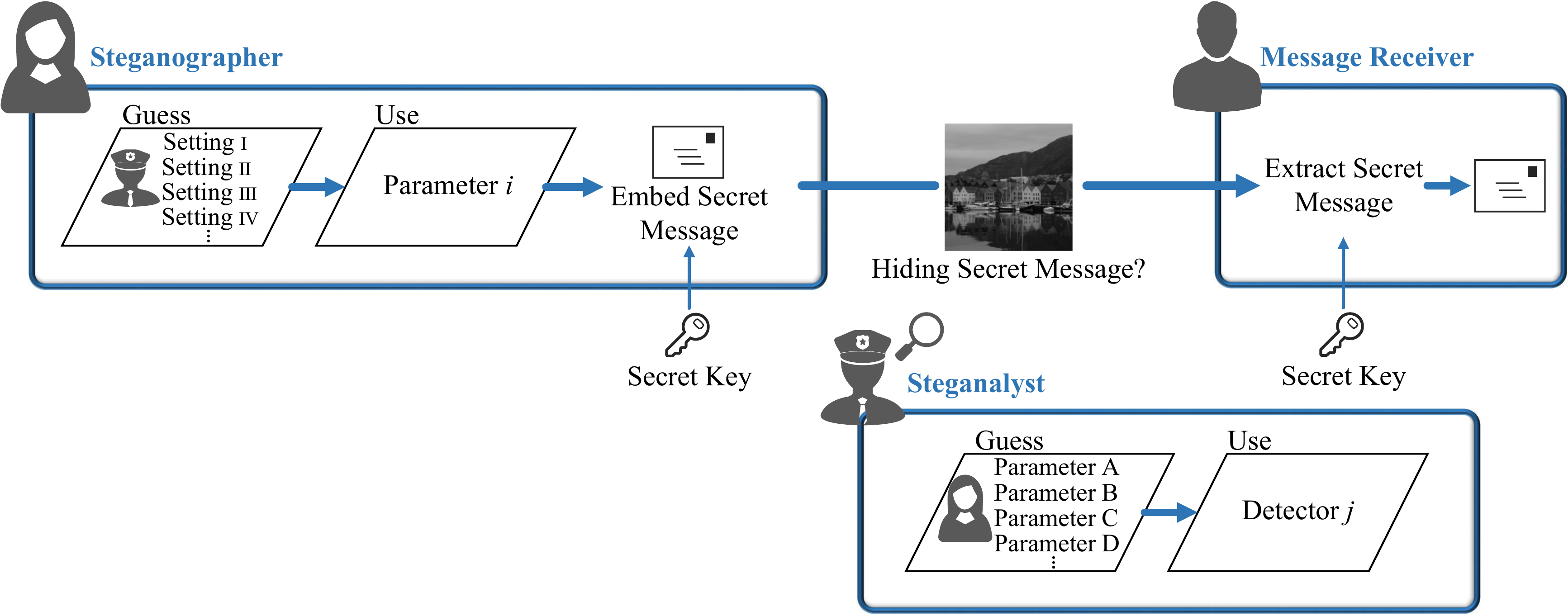}
	\caption{{Scheme of the adversary-aware stego embedding/detection problem considered in this paper. The steganographer sets the internal parameters of its steganographic scheme by assuming that the steganalst will adopt an adversary aware training strategy, but without knowing the setting that he will use to do that. On the other side the steganalyst implements an adversary aware detector by trying to {guess} the internal parameters used by the steganographer.}}
	\label{fig:scheme}
\end{figure*}

Compared to \cite{zhang2018adversarial, Li2018a}, the adversarial stego images generated through the scheme in \cite{Tang2019}  are less detectable by standard
feature based steganalyzers, the rate of modification being only slightly higher than with conventional stego schemes.
When the steganalyst
is aware of the presence of the adversarial steganographer,  robustness against attacks can be improved by adversary-aware training \cite{barni2018adversarial}, that is, by re-training the steganalyzer also with adversarial stego samples.
On the other hand, the steganographer may anticipate the countermeasures adopted by the aware steganalyst and refine the steganographic scheme
to prevent steganalysis.
A problem with this framework is that, arguably, while training the adversary-aware steganalyzer, the steganalyst does not know the exact internal parameters used by the steganograher for adversarial embedding. In the same way, the steganographer does not know the exact kind of images that are used by the steganalyst to train the steganalyzer and hence he can not tune the internal parameters of the steganographic scheme to maximize its deception capability. A pictorial representation of such a dilemma is given in Figure \ref{fig:scheme}.

In this paper, we propose to exit this apparent deadlock by resorting to \textit{game theory} (GT).
In particular, we  introduce a game-theoretic framework according to which each contender of this race of arms, sets the internal parameters of its algorithm trying to strategically anticipate the choice of the other player.

\subsection{Prior Art on Game Theory in Related Security Areas}
The use of game theory to model the interplay between steganography and stenaganalysis has been explored  in several works. In \cite{ker2007batch}, for instance, game theory is used to find the best strategy for a steganographer who can
spread the secret message over several homogeneous cover media (batch steganography),
and for a steganalyst who anticipates this and tries to detect the existence of
at least one secret message (pooled steganalysis). Other interesting game-theoretical
approaches have been proposed recently in the field of content-adaptive steganography.
Content-adaptive steganographic schemes embed the stego-message in the
locations of the cover medium where
the changes are harder to detect \cite{franz2002steganography}.
Sch{\"o}ttle et al. \cite{schottle2012game} have drawn the attention to the fact that, if the
steganalyzer behaves in a strategic manner (and then can recalculate the adaptivity
criterion), adaptive embedding schemes risk to be less secure than random embedding. The authors provide a rigorous approach to secure content-adaptive steganography
by means of a game-theoretic model: the defender and the attacker must decide in
which position to hide and look for evidence of embedding, respectively, by taking
into account the opponent's action. Using the notion of Nash equilibrium, an optimal
adaptive embedding strategy which maximizes the security against a strategic
detector is identified in a simple case. The model has been later extended in \cite{johnson2012hide}.
In \cite{denemark2014detection}, the same approach is applied to the case of a Gaussian cover
and embedding changes based on LSB matching.
Game theory has also been used in many other contiguous security-related fields, e.g., in watermarking \cite{CL2002} and multimedia forensics \cite{Stamm12,Barni2013a}.
A game-theoretic framework to account for the presence of adversaries in general binary detection problems has been studied in \cite{Barni2014,BT18}.
The game-theoretic approach followed in this paper is similar to the one adopted in \cite{abrardo2016game}, where the problem of data fusion in the presence of malicious nodes is studied.

\subsection{Method and Contribution}
In this paper, we assume that the steganographer adopts the adversarial embedding strategy (ADV-EMB) adopted in \cite{Tang2019}, targeting the Xu-net spatial steganalyzer \cite{Xu2016}, while the steganalyst trains an adversary aware version of Xu-net. In doing so, the steganographer must set the parameter $\beta$ of the ADV-EMB algorithm. Such a parameter states the fraction of {\it so called} adjustable elements, that is, those elements (be them pixels or DCT coefficients) that are modified by the algorithm in such a way that the targeted steganalyzer makes a
wrong decision. In \cite{Tang2019}, this parameter is minimized {in order to reduce the embedding distortion}. However, for a given prescribed payload, the steganographer may want to optimize {$\beta$} in such a way to reduce the probability that the stego-image is detected 
by a CNN-based detector. In turn, the steganalyst knows the ADV-EMB scheme adopted by the steganographer and then he trains the detection network {in an adversary-aware fashion}. However, {the steganalyst} does not know the {exact value of the parameter $\beta$ adopted by the} ADV-EMB algorithm and then he has to make an {\textit{educated guess}}. On his side, the steganographer must determine the best $\beta$ without knowing the value used by the CNN {to train the adversary-aware steganalyzer}.
The core of our work is the proposal of a game-theoretic approach to strategically choose the values of $\beta$ used by the steganographer (for embedding) and the steganalyst (for adversary-aware training) \cite{myerson2013game}. By adopting a game-theoretic terminology, the {\it optimum} choice for the two players, represents the equilibrium point of the game (usually, the Nash equilibrium solution is considered \cite{Osb94}), and the evaluation of the performance at the equilibrium permits to assess which contender will gain an advantage over the other in the steganographer vs steganalyst struggle. To the best of our knowledge, this work is the first one that investigates the interplay between CNN-based steganalysis and adversarial attacks from a GT perspective.

\
Throughout the paper, we will refer to the steganalyst as the defender (D) and to the steganographer as the attacker (A).
The main contributions of our work are reported in the following.
\begin{itemize}
\item We formulate a {\textit{non zero-sum game}} \cite{osborne2004introduction} that models the interplay between D, whose goal is to minimize the overall error probability of detecting correctly cover and stego images,  and A, who wants to maximize the probability that the stego images are not detected as such, i.e maximize the missed detection error probability.\footnote{Arguably, the steganographer does not care about the correct or incorrect classification of cover images}.
    \vspace{0.2cm}
    \item
    We prove that, in order to study the equilibria of the game, we  can conveniently solve an associated zero-sum game, where the payoff of the game is defined as the overall error probability, thus greatly simplifying the resolution of the game \footnote{Zero-sum games are generally easier to solve, thanks to the minimax theorem and its relationship with the linear programming duality \cite{neumann1928theorie,adler2013equivalence}}.
\item  We apply the proposed game-theoretic framework to a specific instantiation of the game, when the plain embedding scheme ADV-EMB steganography relies on corresponds to S-UNIWARD \cite{Holub2014}. We derive the optimum strategies for the steganographer and the steganalyst and evaluate the corresponding payoff.
    Eventually, we show experimentally that an improved solution for the staganalyst can be obtained
by considering the distribution of $\beta$ at the equilibrium of the game, and use it to perform aware training.
\end{itemize}

\
The behavior of the optimum strategy for the steganographer confirms
the necessity to find a good trade-off between hiding the adversarial embedding on one hand (by using a low $\beta$), and trying to force the classifier towards a wrong decision on the other (by using a large $\beta$).

\subsection{Organization}

{The rest of this paper is organized as follows.
In Section \ref{sec:basics}, we introduce the notations and basic concepts used in this paper, and present our re-adaptation of the adversarial steganographic scheme where the embedding is controlled by a strength parameter $\beta$.
In Section \ref{sec:game}, we formulate the game between the CNN steganalyzer and the steganographer.
The analysis of the equilibrium solution of the game is carried out in Section \ref{sec:gameeq}.
The methodology of practical equilibrium assessment is detailed in Section \ref{sec:eqpractice}.
In Section \ref{sec:exp}, we experimentally derive and discuss the optimum strategies for the steganalyser and the steganographer and the payoffs at the equilibrium under different settings.
Finally, in Section \ref{sec:conclusion}, we draw some conclusions.}

\section{Technical {Preliminaries}}
\label{sec:basics}

In this section, we introduce the main notations and the metrics used.
We also provide a brief introduction to the ADV-EMB algorithm \cite{Tang2019} and a description of the more general and re-adapted version considered in this paper, where the steganographer can adjust the strength of the attack by choosing the number of modifiable  elements.
%

\subsection{Notation and basic concepts}
\label{sec:basics:notation}

In {the rest of the} paper, we use bold capital letters for matrices {and images},
bold lowercase letters for vectors, and flourish letters for sets.
We let $\mathbf C = (c_{i,j})^{H \times W }$, $\mathbf S = (s_{i,j})^{H \times W }$, and $\mathbf Z^{} =(z_{i,j}^{})^{H \times W }$ be the cover, conventional stego, and adversarial stego images, respectively, where $H$ and $W$ are the height and width of the image.
The sets containing cover, stego, and adversarial stego images are denoted as $\mathcal{C}$, $\mathcal{S}$, and $\mathcal{Z}^{}$, respectively.

Steganalysis can be regarded as a two-class classification problem, where we usually consider stego images as the positive class and cover images as the negative one.
To {build} a CNN classifier, a number of data samples associated with their labels is fed for training in a supervised-learning fashion.
In order to train a CNN classifier, typically, a loss function is defined
and back-propagation is performed according to the gradients of the loss with respective to the learnable parameters/weights.

For the adversary-unaware scenario, we denote a CNN classifier trained on $\mathcal{C}$ and $\mathcal{S}$ as $\phi_{\mathcal{C},\mathcal{S}}$.
Let $\mathbf X$ be an input image and $y$ be its groundtruth label,
where $y=0$ stands for cover and $y=1$ for stego.
The performance of the CNN {are determined} by {the probability of} two kinds of error, {namely the} false alarm probability ($P_{fa}$) and the missed detection probability ($P_{md}$), {defined as:}
\begin{align}\label{}
  P_{fa} & = \Pr \{ \phi_{\mathcal{C},\mathcal{S}}(\mathbf X)
   = 1 | y=0\}  \nonumber\\ & = \Pr \{ \phi_{\mathcal{C},\mathcal{S}}(\mathbf C) = 1 \},
\end{align}
and
\begin{align}\label{}
  P_{md} = & \Pr \{ \phi_{\mathcal{C},\mathcal{S}}(\mathbf X) = 0| y=1 \} \nonumber\\
  = & \Pr \{ \phi_{\mathcal{C},\mathcal{S}}(\mathbf S) = 0\}.
\end{align}

The overall performance can be evaluated by the total error probability ($P_e$), {corresponding to (we assume that the a-priori probabilities of cover and stego images are equal)}:
\begin{equation}\label{}
  P_{e} = \frac{P_{fa}+P_{md} }{2}.
\end{equation}

Let $L(\mathbf X,  y; \phi_{\mathcal{C,S}}^{\mathbf W})$ be the loss function of $\phi_{\mathcal{C,S}}$,
where $\mathbf W = [w_{a,b}]_{(a,b)}$ indicates all the learnable parameters of the CNN.
A typical cross-entropy loss function is defined as
\begin{equation}\label{eq:loss}
  L(\mathbf X,  y; \phi_{\mathcal{C,S}}^{\mathbf W})=
  -y \log( \phi_{\mathcal{C,S}}^{\mathbf W}(\mathbf X) )
  -(1-y) \log( 1-\phi_{\mathcal{C,S}}^{\mathbf W}(\mathbf X) ).
\end{equation}
The parameter $w_{a,b}$ in the $t$-th iteration is updated according to the gradient of the loss function with respect to it,
{i.e.,
$\bigtriangledown_{w_{a,b}} L(\mathbf X, y; \phi_{\mathcal{C,S}}^{\mathbf W})$,}
as follows:
\begin{equation}\label{eq:update}
  w_{a,b}(t) = w_{a,b}(t-1) - \alpha \bigtriangledown_{w_{a,b}} L(\mathbf X, y; \phi_{\mathcal{C,S}}^{\mathbf W}),
\end{equation}
where $\alpha$ is the learning rate.

In a white-box scenario \cite{szegedy2013intriguing}, where the target classifier is assumed to be known {to the} attacker,
an adversarial attack can be launched by modifying the input according to the output loss function in \eqref{eq:loss}
so that the target classifier makes a wrong decision.

One of the core ideas of the adversarial embedding scheme proposed in \cite{Tang2019} is to modify $\mathbf C$ according to both message bits and
the signs of the gradients of the loss function with respective to the input image elements, i.e.,
$\text{sign}( \bigtriangledown_{x_{i,j}} L(\mathbf X, \hat{y}; \phi_{\mathcal{C,S}}))$, where $\hat{y}$ is the false target label.
As the signs of modification and the signs of the gradients are identical,
the output stego image {is} capable of misleading the target classifier.
{However, modifying all elements according to predetermined directions would reduce the actual payload}, leading to a larger distortion compared to conventional embedding.
In addition, in order to mislead the classifier, it may not be necessary to utilize all image elements for adversarial embedding.
Therefore, the image elements are randomly divided into
two groups, where one group is used for adversarial embedding (group of adjustable elements) and the other for conventional embedding.
The fraction $\beta$, indicating the ratio of the adjustable elements over all image elements,
is minimized so to reduce the artifacts introduced in the stego image (under the constraint that the target steganalyzer makes a decision error).
Note that the ADV-EMB scheme can successfully fool the steganalyzer trained by stego
images with conventional embedding.
However, in an adversarial aware steganalysis scenario,
the steganographer does not know the exact kind of images (whether conventional or adversarial stego images)
that are used by the steganalyst to train the steganalyzer, therefore, he may
not use ADV-EMB to maximize its deception capability.

\subsection{Parametric ADV-EMB}
\label{sec:basics:PADV}

We first describe in more detail the ADV-EMB scheme proposed in \cite{Tang2019}. ADV-EMB works under the conventional framework of distortion minimization, in which embedding costs are firstly defined according to
the impact {they have on} each individual image element, and then practical steganographic codes \cite{Filler2011} are employed to minimize
the total distortion associated with the embedding.

Assume $k$ message bits must be embedded in $\mathbf C$.
For a given value $\beta \in [0,1]$, embedding consists of the following steps.
\begin{enumerate}
\item Use a conventional cost function to compute the initial embedding costs of a cover image.
    The resultant costs of positive and negative modifications are respectively denoted as
    $\rho_{i,j}^{+}$ and $\rho_{i,j}^{-}$.
    The cost of no modification $\rho_{i,j}^{0}$ is assumed to be zero.
    \item Randomly select a number of $l_1 = [ H\times W\times (1-\beta) ]$ elements in $\mathbf C$ to form a common group,
    where $[\cdot]$ is the rounding operation.
    The remaining $l_2 = H\times W - l_1$ elements are called adjustable elements and form the adjustable group.
\item Embed $k_1 = [ k\times(1-\beta) ]$ bits into the common group using the initial embedding costs $\{\rho_{i,j}^{+}, \rho_{i,j}^{0}, \rho_{i,j}^{-} \}$ with a distortion minimization coding scheme, such as \cite{Filler2011}.
    The resulting intermediate image is denoted as $\mathbf Z_{c}$.
\item Compute the gradients $\bigtriangledown_{z_{i,j}} L(\mathbf Z_{c}, \hat{y}; \phi_{\mathcal{C},\mathcal{S}})$ of the CNN steganalyzer with respective to image elements using the target label $\hat{y}=0$.
    Update the embedding costs for the adjustable elements as follows:
    \begin{equation}\label{eq:updatecost1}
     \varrho_{i,j}^{+}=
        \begin{cases}
        {\rho_{i,j}^{+}}/{\lambda}, &
        \text{ if $-\bigtriangledown_{z_{i,j}} L(\mathbf Z_{c}, 0; \phi_{\mathcal{C,S}})> 0$},    \\
        {\rho_{i,j}^{+}},  &
        \text{ if $-\bigtriangledown_{z_{i,j}} L(\mathbf Z_{c}, 0; \phi_{\mathcal{C,S}})=0$},    \\
        {\rho_{i,j}^{+}}.{\lambda}, &
        \text{ if $-\bigtriangledown_{z_{i,j}} L(\mathbf Z_{c}, 0; \phi_{\mathcal{C,S}})<0$},    \\
        \end{cases}
      \end{equation}
  \begin{equation}
    \varrho_{i,j}^{0}=\rho_{i,j}^{0},
  \end{equation}
    \begin{equation}\label{eq:updatecost2}
     \varrho_{i,j}^{-}=
        \begin{cases}
        {\rho_{i,j}^{-}}/{\lambda}, &
        \text{ if $-\bigtriangledown_{z_{i,j}} L(\mathbf Z_{c}, 0; \phi_{\mathcal{C,S}})< 0$},    \\
        {\rho_{i,j}^{-}},  &
        \text{ if $-\bigtriangledown_{z_{i,j}} L(\mathbf Z_{c}, 0; \phi_{\mathcal{C,S}})=0$},    \\
        {\rho_{i,j}^{-}}.{\lambda}, &
        \text{ if $-\bigtriangledown_{z_{i,j}} L(\mathbf Z_{c}, 0; \phi_{\mathcal{C,S}})>0$},\\
        \end{cases}
  \end{equation}
where $\lambda=2$ is a scaling factor used to bias the costs towards the desired directions. Embed $k_2= k -k_1$ bits into the adjustable elements by using the updated embedding costs $\{\varrho_{i,j}^{+}, \varrho_{i,j}^{0}, \varrho_{i,j}^{-} \}$  and the same distortion minimization coding scheme used for the common group.
\end{enumerate}

{The above process is applied iteratively, starting from $\beta=0$ and increasing $\beta$ of a small amount $\Delta \beta$ at each iteration. When $\phi_{C,S}(Z) = 0$, i.e., the adversarial stego image $Z$ can fool the steganalyzer,  $Z$ is taken as the output and the embedding process ends, the resulting value of $\beta$ corresponding to the minimum amount of adjustable elements.}


{As we said,
ADV-EMB is effective in fooling the target steganalyzer, but
its effectiveness is reduced when an adversary-aware version of the steganalyzer, trained with adversarial stego images, is considered \cite{Tang2019}. In fact, if the steganalyst is aware of the adversarial embedding strategy adopted by the steganographer, he also knows the parameter $\beta$ to generate the adversarial stego images ($\beta$ is  deterministically derived by solving a minimization problem), and then he can mitigate the effectiveness of adversarial embedding by training an adversary-aware version of the CNN steganalyzer.}

{In this work, we consider a parametric version of the original ADV-EMB scheme, referred to as P-ADV-EMD, where
embedding is implemented without minimizing the amount of adjustable elements, 
but performing Steps 1 to 4, for a given $\beta$.
Let us denote with $\mathbf Z^{\beta}$ the resultant image. The set containing the adversarial stego images attacked with embedding parameter $\beta$ is referred to as $\mathcal{Z}^{\beta}$.}
{Arguably, the steganalyst does not know the value of $\beta$ adopted by the attacker and has to make an educated guess; then, he/she trains the steganalyzer by considering the adversarial embedding performed with the guessed $\beta$.
As confirmed in our experiments (see Section \ref{sec:exp}), when
the steganalyzer trained with a given value of $\beta$ is used to detect adversarial stego images with a mismatched $\beta$,
the detection performance are impaired. The best choice of the parameter $\beta$ to be used by the steganalyst and the steganographer can then be determined in a game-theoretic framework, as stated in the next section.}


\section{CNN-based Adversary-aware Stego Embedding/Detection game.}
\label{sec:game}

Before defining the Adversary-aware Stego Embedding/Detection game, namely the $ASED$ game, in Section \ref{sec:game:SDgame}, we recall some basic concepts of game theory which are necessary to understand the rest the paper.
Then, we will investigate the behavior of the $ASED$ game at the Nash equilibrium.

\subsection{Game theory in a nutshell}
\label{sec:game:GT}

A two-player game is defined by a 4-tuple $(\SS_1,\SS_2,u_1, u_2)$, where $\SS_1 = \{z_{1,1} \dots z_{1,n_1}\}$ and $\SS_2 = \{z_{2,1} \dots z_{2,n_2}\}$ are the set of actions, or strategies, the first and the second player can choose from, and $u_l(z_{1,i}, z_{2,j}), l \in \{1,2\}$, is the payoff of the game for player $l$, when the first player chooses the strategy $z_{1,i}$ $(i \in \{1, \cdots, n_1\})$ and the second chooses $z_{2,j}$  $(j \in \{1, \cdots, n_2\})$. A pair of strategies $(z_{1,i}, z_{2,j})$ is called a profile. In a strategic game,
$\SS_1$, $\SS_2$ and the payoff functions are assumed to be known to the two players, who choose their strategies before starting the game, without knowing the strategy chosen by the other player.

When $u_1(z_{1,i}, z_{2,j}) + u_2(z_{1,i}, z_{2,j}) = 0$, i.e., the players have opposite payoffs, the game is said to be competitive or  {\textit{zero-sum}}. In that case, the payoff of the game can be defined by adopting the perspective of one of the two players.

The goal of game analysis is to determine the existence of equilibrium points, i.e. profiles, that in some way represent a \textit{satisfactory} choice for both players \cite{Osb94}. The most famous  notion of equilibrium is due to {John Nash}. A profile is a Nash equilibrium if no one of the players has any interest in changing his strategy assuming the other does not change his own. For the particular case of a two-player game, a profile $(z_{1,i^*}, z_{2,j^*})$ is a Nash equilibrium if:
\begin{equation}
\begin{array}{ll}
    u_1(z_{1,i^*}, z_{2,j^*}) \ge u_1(z_{1,i}, z_{2,j^*}) & \forall z_{1,i} \in \SS_1,\\
    u_2(z_{1,i^*}, z_{2,j^*}) \ge u_2(z_{1,i^*}, z_{2,j}) & \forall z_{2,j} \in \SS_2.
\end{array}
\label{eq.Nash}
\end{equation}
For a zero-sum game, $u_2 = -u_1$.

The above definition assumes that the players deterministically choose one of the strategies in $\SS_i$ (pure strategy). A more flexible approach consists in letting each player choose a strategy with a certain probability. In this way, we introduce a new game in which the strategies available to the players are probability distributions over $\SS_1$ and $\SS_2$.
The payoffs are redefined in terms of expected payoffs under the probability distributions chosen by the players. A probability distribution ${\bm p}_{l}$ over $\SS_l$, that is,
\begin{equation}
{\bm p}_{l} = \big\{p_{l}(z_{l,i}), z_{l,i} \in \SS_l \big| \sum_i p_{l}(z_{l,i}) = 1, p_{l}(z_{l,i})\ge 0,  \forall i \big\},
\end{equation}
is said a mixed strategy for player $l$.
The definition of Nash equilibrium in mixed strategies extends the one given in \eqref{eq.Nash}. Accordingly, a mixed strategy profile $({\bm p}_{1}^*, {\bm p}_{2}^*)$ is a mixed strategy Nash equilibrium if it satisfies $\bar{u}_1({\bm p}_{1}^*, {\bm p}_{2}^*) \ge \bar{u}_1({\bm p}_{1}, {\bm p}_{2}^*)$ and  $\bar{u}_2({\bm p}_{1}^*, {\bm p}_{2}^*) \ge \bar{u}_2({\bm p}_{1}^*, {\bm p}_{2})$, for any mixed strategy profile ${\bm p}_{1}$ and ${\bm p}_{2}$, where $\bar{u}_l$ denotes the expected payoff for player $l$ under the distribution corresponding to the mixed strategy
profile.
A central result of game theory \cite{Nash50} states that if we allow mixed strategies, then every game with a finite number of players and with a finite number of pure strategies for every player has at least one Nash equilibrium.

\subsection{The $ASED$ game}
\label{sec:game:SDgame}


As we said, we assume that the P-ADV-EMB is implemented  by the steganographer A, by choosing the parameter $\beta$ in a strategic way\footnote{We assume that the steganographer has a perfect knowledge of the target unaware steganalyzer, i.e., the specific CNN architecture adopted and the training set.}. In turn, the steganalyst D has to \textit{guess} the value of $\beta$ used by the steganographer to build the adversary-aware version of the CNN detector.
%
%
As anticipated, we model the interplay between the value of
$\beta$ adopted by A and the one adopted by D as a game.
For sake of clarity, in the following, we indicate with $\beta_A$ the fraction of adjustable elements adopted by A and with $\beta_D$ the value considered by D in
the implementation of the adversary-aware CNN steganalyzer.

%
More formally, A selects  $\phi_{\mathcal{C},\mathcal{S}}$, i.e., the unaware version of the steganalyzer, as the target steganalyzer, 
and the adversarial stego images are generated by considering $\beta_A$. On the other hand, D selects $\beta_D$ and considers $\phi_{\mathcal{C},\mathcal{Z}^{\tiny \beta_{D}}}$ for detection. With these
ideas in mind, we are now ready to define the CNN-based $ASED$ game.

\begin{defn}
\label{def:SDgame}
The $ASED(\mathcal{S}_{A}, \mathcal{S}_{D}, u_A, u_D)$ game is a two-player,
non-zero-sum, strategic game played by the steganalyst (D) and the steganographer (A), defined by the
following strategies and payoffs.
\begin{itemize}
\item{The sets of strategies  the steganographer (A) and  the steganalyst (D) can choose from are, respectively, the set of possible values of $\beta_A$ and $\beta_D$:
\begin{equation}
\begin{aligned}
\mathcal{S}_A &= \{ \beta_A \in [0,1]\}, \\
\mathcal{S}_D &= \{ \beta_D \in [0,1]\}.
\label{eq.DFgameS}
\end{aligned}
\end{equation}
}
\item{The payoff of the steganalyst (D) is defined as the negative error probability of the CNN classifier;
that is
\begin{align}
u_D(\beta_A, \beta_D) =  - P_e(\beta_A, \beta_D) = - \frac{1}{2}(P_{fa}(\beta_D) + P_{md}(\beta_A, \beta_D))
\label{eq.Pe}
\end{align}
where
\begin{equation}
P_{fa}(\beta_A, \beta_D) = \Pr\{\phi_{\mathcal{C},\mathcal{Z}^{\tiny \beta_{D}}}(\mathbf C) = 1\},
\end{equation}
and
\begin{equation}
P_{md}(\beta_A, \beta_D) = \Pr\{\phi_{\mathcal{C},\mathcal{Z}^{\tiny \beta_{D}}}(\mathbf Z^{\beta_A}) = 0\}.
\end{equation}
}
\item{The payoff of the steganographer (A)  is defined as the missed detection probability, i.e., $u_A(\beta_A, \beta_D)= P_{md}(\beta_A, \beta_D)$.
}
\end{itemize}
\end{defn}

{We stress that the above non zero-sum game formulation is a novelty with respect to prior art in the field, where zero sum game formulations have been considered; moreover, it better models the general adversarial stego detection scenario, where the goal of the adversary is to conceal the presence of the message in the stego images, that is, to pass off a stego as a cover, and not to induce general misclassification error.}

In the above definition, the sets
of strategies available to A and D are
continuous sets.  However, to derive
the equilibrium point for the $ASED$ game,
we will consider discrete sets of strategies by properly quantizing the values of $\beta_A$ and $\beta_D$. The quantized sets of strategies are indicated by $\mathcal{S}_A^q$ and $\mathcal{S}_D^q$. Then, we consider the pair of strategies $(\beta_A,\beta_D) \in \mathcal{S}_{A}^q \times \mathcal{S}_{D}^q$.
We denote with  $(- {\mathbf P_{e}})$, res. ${\mathbf P_{md}}$, the payoff matrices of  D, res. A, where $\mathbf P_{e} = \left[P_e(\beta_A, \beta_D)\right]_{\beta_A \in \mathcal{S}_{A}^q, \beta_D \in \mathcal{S}_{D}^q}$, $\mathbf P_{md} = \left[P_{md}(\beta_A, \beta_D)\right]_{\beta_A \in \mathcal{S}_{A}^q, \beta_D \in \mathcal{S}_{D}^q}$.

In the sequel, we will always consider the $ASED$ game with quantized sets of strategies, namely $ASED(\mathcal{S}_{A}^q, \mathcal{S}_{D}^q, u_A, u_D)$, unless stated differently.

\section{Equilibrium Point Analysis (of the $ASED$ Game)}
\label{sec:gameeq}

{The ultimate goal of our analysis is to} determine the equilibrium point(s) of the quantized version of the $ASED$ game,
which, as we will see, can be found in mixed strategies.


Let $\bm{p}_D$, res. $\bm{p}_A$, indicate the mixed strategies vectors, that is, the (column) vectors with the probability distribution over the possible values of $\beta_D$, res. $\beta_A$, in $\mathcal{S}_{D}^q$ and $\mathcal{S}_{A}^q$. For a given mixed strategy profile $(\bm{p}_A,\bm{p}_D)$, the expected payoffs of A and D can be computed as:
\begin{align}
\label{avg_payoff}
\bar{u}_A(\bm{p}_A,\bm{p}_D) = \sum_{\beta_A \in \mathcal{S}_{A}^q} p_A(\beta_A)  \sum_{\beta_D \in \mathcal{S}_{D}^q}   P_{md} (\beta_A, \beta_D)  p_D(\beta_D),\\
\label{avg_payoff2}
\bar{u}_D(\bm{p}_A,\bm{p}_D) = - \sum_{\beta_A \in \mathcal{S}_{A}^q} p_A(\beta_A)  \sum_{\beta_D \in \mathcal{S}_{D}^q}   P_{e} (\beta_A, \beta_D)  p_D(\beta_D).
\end{align}

A mixed strategies Nash equilibrium profile $(\bm{p}_A^*, \bm{p}_D^*)$ is a pair of mixed strategies for which we have:
%
\begin{align}
& {(\bm{p}_A^*)}^T \cdot  {\mathbf P_{md}}\cdot  \bm{p}_D^*  = \max_{\bm{p}_A}\bigg\{ {(\bm{p}_A)}^T \cdot {\mathbf P_{md}}\cdot  \bm{p}_D^* \bigg| \sum_{\beta_A \in \mathcal{S}_A^q} p_A(\beta_A) = 1, p_A(\beta_A) \ge 0 \bigg\}, \nonumber\\
& {(\bm{p}_A^*)}^T \cdot  {\mathbf P_{e}}\cdot  \bm{p}_D^*  = \min_{\bm{p}_D}\bigg\{ {(\bm{p}_A^*)}^T \cdot {\mathbf P_{e}}\cdot  \bm{p}_D \bigg| \sum_{\beta_D \in \mathcal{S}_D^q} p_D(\beta_D) = 1,  p_D(\beta_D) \ge 0 \bigg\}.
\label{eq.NashNonZS}
\end{align}
%
%

In general, solving a non-zero-sum game, i.e. finding the Nash equilibrium (equilibria) of the game, is not easy \cite{daskalakis2009complexity}.
%
In hindsight, in our case, the problem
 can be simplified by observing that the $P_{fa}$ corresponding to the steganalyzer $\phi_{\mathcal{C},\mathcal{Z}^{\tiny \beta_{D}}}$
does not depend on $\beta_A$. Specifically, we can write the following equivalence:
\begin{equation}
 {(\bm{p}_A)}^T \cdot  {\mathbf P_{e}} \cdot  \bm{p}_D = \frac{1}{2}\left\{{(\bm{p}_A)}^T \cdot  {\mathbf P_{md}} \cdot  \bm{p}_D +  \sum_{\beta_D \in \mathcal{S}_D^q} P_{fa}(\beta_D) p_D(\beta_D)\right\}.
\end{equation}
The above relation follows immediately by observing that
\begin{equation}
 {(\bm{p}_A)}^T \cdot  {\mathbf P_{e}} \cdot  \bm{p}_D  = \frac{1}{2} \left\{ {(\bm{p}_A)}^T \cdot  {\mathbf P_{md}} \cdot  \bm{p}_D +
 {(\bm{p}_A)}^T {\mathbf P_{fa}}  \cdot  \bm{p}_D \right\} \nonumber
\end{equation}
where $\mathbf P_{fa} = \left[P_{fa}(\beta_A, \beta_D)\right]_{\beta_A \in \mathcal{S}_{A}^q, \beta_D \in \mathcal{S}_{D}^q}$, which is constant over the rows. Then:
\begin{equation}
\begin{aligned}
 {(\bm{p}_A)}^T {\mathbf P_{fa}}  \cdot  \bm{p}_D  = & \sum_{\beta_D \in \mathcal{S}_D^q}\left(\sum_{\beta_A \in \mathcal{S}_A^q} P_{fa}(\beta_A, \beta_D)\cdot p_A(\beta_A)\right) \cdot p_D(\beta_D) \\
  & = \sum_{\beta_D \in \mathcal{S}_D^q} P_{fa}(\beta_D) \cdot \left(\sum_{\beta_A \in \mathcal{S}_A^q} p_A(\beta_A)\right) \cdot p_D(\beta_D)\\
  & = \sum_{\beta_D \in \mathcal{S}_D^q} P_{fa}(\beta_D)\cdot p_D(\beta_D).
\end{aligned}
\end{equation}
%
%
Accordingly, the mixed strategy $\bm{p}_A^*$ at the Nash equilibrium can  be equivalently obtained by solving the maximization below:
\begin{equation}
\label{equivalentMax}
\bm{p}_A^* = \arg\max_{\bm{p}_A} \bigg\{ {(\bm{p}_A)}^T \cdot {\mathbf P_{e}} \cdot  \bm{p}_D^* \bigg| \sum_{\beta_A \in \mathcal{S}_A^q} p_A(\beta_A) = 1, p_A > 0 \bigg\},
\end{equation}
where ${\mathbf P_{e}}$ is considered as payoff matrix for A.

Given the above derivation, the following property holds immediately.\footnote{The property can also be proved with reference to the continuous game $ASED(\mathcal{S}_{A}, \mathcal{S}_{D}, u_A, u_D)$ (the derivation is obtained by replaing the sums with integrals in equations \eqref{eq.NashNonZS} through \eqref{equivalentMax}).
}

\begin{theorem}
\label{property}
A profile  $(\bm{p}_A^*, \bm{p}_D^*)$ is a Nash equilibrium of the non-zero-sum strategic game $AESD(\mathcal{S}_{A}^q, \mathcal{S}_{D}^q, u_A, u_D)$, if and only if it is a Nash equilibrium of the \textit{zero-sum} strategic game $ASED'(\mathcal{S}_{A}^q, \mathcal{S}_{D}^q, u)$
where $u = {\mathbf P_{e}}$ is the payoff of the game, defined by adopting the steganalyst's  perspective (i.e., A aims at maximizing $u$, while D aims at minimizing it).
\end{theorem}

Thanks to Property \ref{property}, in order to find the equilibrium point of the $ASED$ game, we can conveniently solve the zero-sum game $ASED'$.
This represents a  great advantage, since finding  the Nash equilibrium of a zero-sum game is significantly easier.
In particular, since the set of strategies is discrete and finite, {the equilibrium point} can be derived by relying on the minimax theorem \cite{Osb94}. According to such a theorem,
the mixed strategies Nash equilibrium $(\bm{p}_A^*, \bm{p}_D^*)$ can be obtained by solving separately the following max-min and min-max problems:
\begin{equation}
\begin{aligned}
& \bm{p}_A^* = {\arg}\max_{\bm{p}_A(\mathcal{S}_{A}^q)} \min_{\bm{p}_{D}(\mathcal{S}_{D}^q)} \bm{p}_A^T  \cdot {\mathbf P_{e}} \cdot  \bm{p}_{D} \\
& \bm{p}_{D}^* = {\arg}\min_{\bm{p}_{D}(\mathcal{S}_{D}^q)}  \max_{\bm{p}_A(\mathcal{S}_A^q)}  \bm{p}_A^T \cdot {\mathbf P_{e}}  \cdot   \bm{p}_{D},
\label{eq.MinMax}
\end{aligned}
\end{equation}
where the $\max$ and $\min$
are restricted to the set of probability vectors, that is, the vectors for which
$\{\bm{p}_A: \sum_{\beta_A \in \mathcal{S}_{A}^q} p_A(\beta_A) = 1, p_A(\beta_A) \ge 0, \forall \beta_A\}$ and $\{\bm{p}_D:  \sum_{\beta_D \in \mathcal{S}_{D}^q} p_D(\beta_D) = 1, p_D(\beta_D) > 0, \forall \beta_D\}$.
The above system can be reduced to the solution of a linear programming (LP) problem, see \cite{charnes1957management}.
The expected payoffs of D and A at the equilibrium of the non-zero-sum game are $\bar{u}_D(\bm{p}_A^*, \bm{p}_D^*)$ and $\bar{u}_A(\bm{p}_A^*, \bm{p}_D^*)$, computed as in Equation \eqref{avg_payoff} and \eqref{avg_payoff2}.




\section{Equilibrium Assessment in Practice}
\label{sec:eqpractice}

Having established that solving the $ASED'$ game is equivalent to finding the equilibrium points of the $ASED$ game, our next step is to investigate the behavior of the $ASED$ game in a practical scenario and analyze the achievable performance of the steganalyst and the steganographer, when they adopt CNN aware training and the P-ADV-EMB scheme respectively, and tune the values of $\beta_D$ and $\beta_A$ strategically. Specifically, the goal of our research is to
study the equilibrium point of the game in pure or mixed
strategies, and analyze the behavior and the payoff of the steganographer and
the steganalyst at the equilibrium.
The performance at the equilibrium are then compared to those obtained by adopting a worst case approach and those achieved by training the steganalyzer with a proper mixture of adversarial stego samples obtained by adopting different values of $\beta_D$.
%


\subsection{Experimental setting}
\label{sec:methodology:SET}

The setup we have used to conduct our experiments is described in the following.
We {applied} the P-ADV-EMB algorithm in the spatial domain
by selecting S-UNIWARD \cite{Holub2014} as the baseline scheme for conventional embedding.
Xu-net \cite{Xu2016} was used as the CNN classifier for its fast convergence under a moderate size of the training set.
The {Xu-net} steganalyzer trained with conventional S-UNIWARD stego images, i.e., $\phi_{\mathcal{C}_{trn},\mathcal{S}_{trn}}$ (or, equivalently,  $\phi_{\mathcal{C}_{trn},\mathcal{Z}^{0}_{trn}}$), was used as the target steganalyzer used to generate the adversarial stego images $\mathbf Z^{\beta_A}$.

For the evaluation, we used the BOSSBase v1.01 \cite{Bas2011} dataset, which contains 10,000 cover images of size $512\times 512$.
The payload was set to 0.4 bpp (bit per pixel).
In the experiments, firstly, we quantized $\beta$ based on the considerations we made in Section \ref{sec:methodology:BETA}.
Then, for a given $\beta$, we randomly split 10,000 pairs of cover images and their corresponding stego images into three disjoint subsets, i.e., training set $\{\mathcal{C}_{trn},\mathcal{Z}_{trn}^{\beta}\}$, validation set $\{\mathcal{C}_{val},\mathcal{Z}_{val}^{\beta}\}$, and test set $\{\mathcal{C}_{tst},\mathcal{Z}_{tst}^{\beta}\}$, with 4000, 1000, and 5000 pairs of images, respectively.
The training set was employed to train the learnable parameters/weights in the CNN steganalyzer, while the validation set was used for the selection of the best-performing model, i.e., the best-performing parameters/weights in the CNN, to prevent over-fitting to the training set.
The reported performance of the CNN steganalyzer was evaluated on the test sets.
Finally, to build the payoff matrices $\mathbf{P}_e$ and $\mathbf{P}_{md}$, each CNN steganalyzer $\phi_{\mathcal{C}_{trn},\mathcal{Z}^{\beta_D}_{trn}}$, $\beta_D \in \mathcal{S}_D^q$, was run on the set $\{\mathcal{C}_{tst},\mathcal{Z}_{tst}^{\beta_A}\}$ for all $\beta_A \in \mathcal{S}_A^q$.

CNN training and testing was performed on TensorFlow with Python interface and a NVIDIA Tesla P100 GPU card.
The weights of CNN convolutional filters and fully connected layers were initialized by using a normal distribution with zero mean and standard deviation {equal} to 0.01. With regard to learning, we used stochastic gradient descent, with momentum {equal to} 0.9 and initial learning rate 0.001.
The learning rate decay was set to 90\% every 5,000 training steps.
The batch size in each iteration was set to 50 (25 cover/stego pairs).
The training stage {lasted} 110,000 iterations, and validation was performed every 5,000 iterations.


\subsection{Investigated solutions}
\label{sec:methodology:SOLUTION}

\subsubsection{Nash equilibrium}

To stat with, we measured the performance at the equilibrium of the stego detection game by following the approach presented in Section \ref{sec:gameeq}.
Specifically, by Property \ref{property}, we obtained the Nash equilibrium profile $(\bm{p}_A^*, \bm{p}_D^*)$ by solving the LP problem associated to the zero-sum game formulation $ASED'$, with payoff matrix ${\mathbf P_{e}}$ having entries  $|\mathcal{S}_A^q| \times |\mathcal{S}_D^q|$.
Then, we computed the payoffs of D and A according to Equations \ref{avg_payoff} and \ref{avg_payoff2}.
We used the linear programming tools of Matlab Optimization Toolbox \cite{Toolbox2016} to solve the LP problem by means of the simplex algorithm.

\subsubsection{Worst case solution}
\label{sec:::worstcase}

As a second possibility, we considered the case in which A and B adopts a conservative worst case approach.
In this case, A and D choose the strategy which maximizes their own payoff in the worst case with respect to the move of the opponent \footnote{Note that, in a zero-sum game, this strategy may not correspond to the strategy that maximizes the opponent's payoff.}.

{In our setup}, the worst case strategy for A corresponds to the {value of $\beta_A$}  that maximizes the missed detection probability when D plays the  strategy that minimizes it, that is, given the payoff matrix $\mathbf P_{md}$,
\begin{equation}\label{eq:worstcaseA}
  \hat{\beta}_A =  \arg {\max \limits_{\beta_{A} \in \mathcal{S}^q_A}} \left\{ {\min \limits_{\beta_{D} \in \mathcal{S}^q_D}} P_{md}(\beta_A, \beta_D) \right\}.
\end{equation}
The corresponding worst case solution is  $\min_{\beta_{D}} P_{md}(\hat{\beta}_A, \beta_D)$.

Likewise, given  D's payoff matrix $-\mathbf P_{e}$,  the worst case strategy for D is
\begin{equation}\label{eq:worstcaseD}
   \hat{\beta}_D = \arg {\min \limits_{\beta_{D} \in \mathcal{S}^q_D}} \left\{ {\max \limits_{\beta_{A} \in \mathcal{S}^q_A}} P_e(\beta_A, \beta_D) \right\}.
\end{equation}
Usually, $\hat{\beta}_D$ (res. $\hat{\beta}_A$) is different from the value obtained from the inner minimization in \eqref{eq:worstcaseA}  (res. in \eqref{eq:worstcaseD}). When the solution of Equations \eqref{eq:worstcaseA} and \eqref{eq:worstcaseD} is the same,
the profile $(\hat{\beta}_A, \hat{\beta}_D)$ corresponds to a pure strategy Nash equilibrium.

In the next section, we compare the worst case solution to the Nash equilibrium of the game, to show the advantage provided by the game-theoretic analysis.

\subsubsection{Training over a mixture of strategies}

Eventually, we considered the solution obtained by training the CNN on a mixture of $\beta$ values. In particular, we considered the mixture corresponding to the equilibrium mixed strategy for D, namely $\bm{p}_D^* = [p_D^*(\beta_D)]_{\beta_D \in \mathcal{S}_D^q}$,
to build the adversarial stego image set $\mathcal{Z}^{\mathbfit{p}_D^*}$, where adversarial embedding was performed with various $\beta_D \in \mathcal{S}^q_D$ according to the probability vector $\bm{p}_D^*$. The trained model is denoted with $\phi_{\mathcal{C}_{trn},\mathcal{Z}^{\mathbfit{p}_D^*}_{trn}}$ in this case.
In practice, we exploited the equilibrium solution of the game from the perspective of the steganalyst, thus going beyond a strictly game-theoretic analysis.
More specifically, by training the CNN with a mixture of $\beta$, we gave an advantage to the steganalyst, since the steganographer is assumed to keep playing at the equilibrium of the game, which was found without considering the possibility that the steganalyzer was trained on a mixture of $\beta$'s.\footnote{Training on mixtures of $\beta$'s was not included in the set of strategies of the game.}
We also considered the case in which adversarial embedding was performed by considering an uniform distribution of $\beta_D$ across the training set
$\mathcal{S}^q_D$. In this case, the trained steganalyzer is denoted by $\phi_{\mathcal{C}_{trn},\mathcal{Z}^{\text{uni}}_{trn}}$.
%


\subsection{Quantization of $\beta$ values}
\label{sec:methodology:BETA}

In order to approximate the behavior of the continuous game  (see Definition \ref{def:SDgame}), we should consider a fine enough quantization of  $\beta_A$ and $\beta_D$. However, considering a very fine quantization increases {dramatically} the computational burden of the tests, since a CNN model has to be trained for every value of $\beta_D \in \mathcal{S}^q_D$.
In our experiments, we first quantized $\beta_A$ and $\beta_D$  with a uniform quantization step size of 0.05,
then we considered non-uniform quantization steps with a smaller step size in the region of $(\beta_A,\beta_D)$ values where the payoff varies more rapidly. 
We considered only the case of equal quantization strategies for D and A, that is $\mathcal{S}_A^q \equiv \mathcal{S}_D^q \equiv \mathcal{S}^q$.

\begin{table*}
\caption{$P_{e}$ (in\%) with a uniform quantization step size of 0.05 on $\beta$.
(Wave: the case of conventional embedding and detection.
Bold: the case when the steganalyst has a worse performance compared to the case of conventional embedding and detection.
Shade: the case when the steganalyst uses a matched parameter for detection, i.e., $\beta_D=\beta_A$.
Dash underline: the case when the steganalyst has a better performance with a mismatched parameter than with a matched parameter.
Underline: worst case solution for steganalyst.)}

\centering %
\label{tab:betaU_pe}       
\setlength{\tabcolsep}{0.35mm}{
\begin{tabular}{cccccccccccccccccccccc}
\toprule
$\beta_{A} \verb|\| \beta_{D}$ & 0.00 & 0.05 & 0.10 & 0.15 & 0.20 & 0.25 & 0.30 & 0.35 & 0.40 & 0.45 & 0.50 & 0.55 & 0.60 & 0.65 & 0.70 & 0.75 & 0.80 & 0.85 & 0.90 & 0.95 & 1.00 \\
\noalign{\smallskip}\hline\noalign{\smallskip}
0.00 & \cellcolor[HTML]{C0C0C0}{\uwave{\textbf{20.9}}} & \textbf{26.4} & \textbf{50.5} & \textbf{50.3} & \textbf{50.1} & \textbf{50.1} & \textbf{50.1} & \textbf{50.1} & \textbf{50.0} & \textbf{50.0} & \textbf{50.0} & \textbf{50.0} & \textbf{50.0} & \textbf{50.0} & \textbf{50.0} & \textbf{50.0} & \textbf{50.0} & \textbf{50.0} & \textbf{50.0} & \textbf{50.0} & \textbf{50.0} \\
0.05 & \textbf{39.0} & \cellcolor[HTML]{C0C0C0}\textbf{26.7} & \textbf{42.7} & \textbf{48.5} & \textbf{49.5} & \textbf{49.8} & \textbf{49.7} & \textbf{49.7} & \textbf{49.8} & \textbf{50.0} & \textbf{49.9} & \textbf{50.0} & \textbf{50.0} & \textbf{50.0} & \textbf{50.0} & \textbf{50.0} & \textbf{50.0} & \textbf{50.0} & \textbf{50.0} & \textbf{50.0} & \textbf{50.0} \\
0.10 & \textbf{54.7} & \textbf{27.1} & \cellcolor[HTML]{C0C0C0}20.1 & \textbf{30.9} & \textbf{45.0} & \textbf{47.2} & \textbf{47.1} & \textbf{48.1} & \textbf{48.7} & \textbf{49.5} & \textbf{49.6} & \textbf{49.8} & \textbf{49.9} & \textbf{49.8} & \textbf{49.9} & \textbf{49.8} & \textbf{49.9} & \textbf{49.9} & \textbf{49.9} & \textbf{49.8} & \textbf{49.9} \\
0.15 & \textbf{57.0} & \textbf{28.2} & {\dashuline{8.0}} & \cellcolor[HTML]{C0C0C0}10.1 & \textbf{24.6} & \textbf{29.1} & \textbf{34.2} & \textbf{42.8} & \textbf{43.6} & \textbf{47.5} & \textbf{47.8} & \textbf{48.3} & \textbf{49.1} & \textbf{48.9} & \textbf{49.3} & \textbf{49.0} & \textbf{49.4} & \textbf{48.3} & \textbf{48.8} & \textbf{48.8} & \textbf{49.2} \\
0.20 & \textbf{57.7} & \textbf{28.8} & {\dashuline{4.8}} & {\dashuline{4.3}} & \cellcolor[HTML]{C0C0C0}6.9 & 8.6 & 12.7 & \textbf{26.1} & \textbf{22.3} & \textbf{36.4} & \textbf{37.1} & \textbf{38.8} & \textbf{42.1} & \textbf{44.9} & \textbf{43.9} & \textbf{42.5} & \textbf{45.3} & \textbf{42.7} & \textbf{44.2} & \textbf{45.8} & \textbf{46.7} \\
0.25 & \textbf{57.9} & \textbf{30.1} & 3.7 & {\dashuline{2.6}} & {\dashuline{2.9}} & \cellcolor[HTML]{C0C0C0}3.1 & 4.6 & 7.4 & 6.9 & 14.4 & 16.3 & 19.1 & \textbf{24.1} & \textbf{29.3} & \textbf{31.3} & \textbf{29.1} & \textbf{35.8} & \textbf{36.0} & \textbf{38.0} & \textbf{41.1} & \textbf{43.0} \\
0.30 & \textbf{58.1} & \textbf{31.4} & 3.0 & {\dashuline{1.9}} & {\dashuline{1.6}} & {\dashuline{1.8}} & \cellcolor[HTML]{C0C0C0}2.0 & 2.5 & 2.0 & 5.2 & 5.5 & 7.5 & 10.7 & 14.3 & 18.9 & 17.8 & \textbf{24.4} & \textbf{28.0} & \textbf{30.3} & \textbf{34.8} & \textbf{38.0} \\
0.35 & \textbf{58.2} & \textbf{32.5} & 2.7 & 1.4 & 1.1 & 1.1 & 1.2 & \cellcolor[HTML]{C0C0C0}1.1 & 0.6 & 2.6 & 2.0 & 2.9 & 4.7 & 6.6 & 10.0 & 10.0 & 13.5 & 19.1 & \textbf{21.7} & \textbf{26.7} & \textbf{31.2} \\
0.40 & \textbf{58.3} & \textbf{33.9} & 2.5 & 1.1 & 0.9 & 0.8 & 0.9 & 0.9 & \cellcolor[HTML]{C0C0C0}0.3 & 1.0 & 0.6 & 1.2 & 1.8 & 2.8 & 4.8 & 4.6 & 6.5 & 10.7 & 13.4 & 18.0 & \textbf{23.4} \\
0.45 & \textbf{58.4} & \textbf{34.8} & 2.4 & 1.0 & 0.8 & 0.6 & 0.7 & 0.7 & {\dashuline{0.2}} & \cellcolor[HTML]{C0C0C0}0.6 & 0.3 & 0.5 & 0.8 & 1.3 & 2.1 & 2.1 & 3.0 & 5.3 & 7.1 & 9.7 & 15.4 \\
0.50 & \textbf{58.5} & \textbf{35.5} & 2.3 & 1.0 & 0.7 & 0.4 & 0.5 & 0.7 & 0.2 & 0.4 & \cellcolor[HTML]{C0C0C0}0.2 & 0.3 & 0.4 & 0.8 & 1.0 & 1.0 & 1.4 & 2.5 & 3.5 & 4.6 & 9.0 \\
0.55 & \textbf{58.6} & \textbf{36.2} & 2.2 & 0.9 & 0.7 & 0.5 & 0.5 & 0.6 & 0.2 & 0.3 & 0.2 & \cellcolor[HTML]{C0C0C0}0.2 & 0.2 & 0.6 & 0.5 & 0.5 & 0.8 & 1.1 & 1.8 & 2.0 & 5.1 \\
0.60 & \textbf{58.6} & \textbf{36.7} & 2.2 & 0.9 & 0.7 & 0.4 & 0.4 & 0.6 & 0.2 & 0.2 & {\dashuline{0.1}} & {\dashuline{0.1}} & \cellcolor[HTML]{C0C0C0}0.2 & 0.5 & 0.3 & 0.3 & 0.4 & 0.6 & 0.8 & 0.9 & 2.9 \\
0.65 & \textbf{58.6} & \textbf{37.4} & 2.2 & 0.8 & 0.6 & 0.3 & 0.3 & 0.6 & 0.2 & 0.2 & {\dashuline{0.1}} & {\dashuline{0.1}} & {\dashuline{0.1}} & \cellcolor[HTML]{C0C0C0}0.3 & 0.2 & 0.3 & 0.3 & 0.3 & 0.5 & 0.5 & 1.6 \\
0.70 & \textbf{58.6} & \textbf{37.6} & 2.1 & 0.8 & 0.6 & 0.3 & 0.3 & 0.6 & 0.2 & 0.2 & {\dashuline{0.1}} & {\dashuline{0.1}} & {\dashuline{0.1}} & 0.3 & \cellcolor[HTML]{C0C0C0}0.2 & 0.2 & 0.2 & 0.2 & 0.4 & 0.3 & 1.1 \\
0.75 & \textbf{58.7} &
\uline{\textbf{37.8}} & 2.1 & 0.8 & 0.6 & 0.3 & 0.3 & 0.6 & 0.2 & 0.2 & {\dashuline{0.1}} & {\dashuline{0.1}} & {\dashuline{0.1}} & 0.3 & {\dashuline{0.1}} & \cellcolor[HTML]{C0C0C0}0.2 & 0.2 & 0.2 & 0.2 & 0.2 & 0.8 \\
0.80 & \textbf{58.7} & \textbf{37.7} & 2.1 & 0.8 & 0.6 & 0.3 & 0.3 & 0.6 & 0.2 & 0.2 & 0.2 & 0.2 & {0.1} & 0.3 & { 0.1} & 0.2 & \cellcolor[HTML]{C0C0C0}0.1 & 0.1 & 0.1 & 0.2 & 0.6 \\
0.85 & \textbf{58.7} & \textbf{37.7} & 2.1 & 0.8 & 0.6 & 0.4 & 0.4 & 0.6 & 0.2 & 0.3 & 0.2 & 0.2 & 0.2 & 0.3 & 0.2 & 0.2 & 0.1 & \cellcolor[HTML]{C0C0C0}0.1 & 0.1 & 0.2 & 0.4 \\
0.90 & \textbf{58.7} & \textbf{37.4} & 2.1 & 0.8 & 0.6 & 0.4 & 0.4 & 0.7 & 0.2 & 0.4 & 0.3 & 0.3 & 0.3 & 0.5 & 0.2 & 0.2 & 0.1 & 0.1 & \cellcolor[HTML]{C0C0C0}0.1 & 0.1 & 0.3 \\
0.95 & \textbf{58.7} & \textbf{37.5} & 2.1 & 0.8 & 0.7 & 0.4 & 0.5 & 0.8 & 0.3 & 0.5 & 0.5 & 0.5 & 0.4 & 0.6 & 0.4 & 0.3 & 0.2 & 0.2 & {\dashuline{0.1}} & \cellcolor[HTML]{C0C0C0}0.2 & 0.3 \\
1.00 & \textbf{58.7} & \textbf{37.2} & 2.1 & 0.8 & 0.7 & 0.5 & 0.7 & 0.9 & 0.5 & 0.6 & 0.6 & 0.6 & 0.6 & 0.8 & 0.6 & 0.5 & 0.2 & 0.2 & 0.2 & 0.2 & \cellcolor[HTML]{C0C0C0}0.2 \\
\bottomrule
\end{tabular}}
\end{table*}
%
%

\begin{table*}
\vspace{0.2cm}
\caption{$P_{md}$ (in\%) with a uniform quantization step size of 0.05 on $\beta$.
(Wave: the case of conventional embedding and detection.
Shade: the case when the steganalyst uses a matched parameter for detection, i.e., $\beta_D=\beta_A$.
Bold: the case when the steganographer has a better performance compared to the case of conventional embedding and detection.
Double underline: worst case solution for steganographer.)}
\centering %
\label{tab:betaU_pmd}       
\setlength{\tabcolsep}{0.33mm}{
\begin{tabular}{cccccccccccccccccccccc}
\toprule
$\beta_{A} \verb|\| \beta_{D}$ & 0.00 & 0.05 & 0.10 & 0.15 & 0.20 & 0.25 & 0.30 & 0.35 & 0.40 & 0.45 & 0.50 & 0.55 & 0.60 & 0.65 & 0.70 & 0.75 & 0.80 & 0.85 & 0.90 & 0.95 & 1.00 \\
\noalign{\smallskip}\hline\noalign{\smallskip}
0.00 & \cellcolor[HTML]{C0C0C0}\uwave{24.0} & 29.8 & 96.8 & 99.2 & 99.1 & 99.6 & 99.5 & 99.1 & 99.8 & 99.7 & 100.0 & 99.9 & 99.9 & 99.6 & 99.9 & 99.8 & 99.9 & 100.0 & 99.9 & 100.0 & 99.9 \\
0.05 & 60.3 & \cellcolor[HTML]{C0C0C0}30.6 & 81.3 & 95.6 & 97.8 & 98.9 & 98.8 & 98.2 & 99.3 & 99.6 & 99.8 & 99.8 & 99.8 & 99.5 & 99.9 & 99.7 & 99.9 & 99.9 & 99.8 & 99.8 & 99.9 \\
0.10 & 91.7 & \uuline{31.2} & \cellcolor[HTML]{C0C0C0}36.1 & 60.4 & 88.8 & 93.7 & 93.6 & 95.0 & 97.1 & 98.7 & 99.2 & 99.4 & 99.6 & 99.1 & 99.7 & 99.4 & 99.7 & 99.7 & 99.6 & 99.6 & 99.7 \\
0.15 & 96.3 & 33.5 & \textbf{11.8} & \cellcolor[HTML]{C0C0C0}\textbf{18.7} & 48.0 & 57.5 & 67.8 & 84.4 & 86.8 & 94.7 & 95.5 & 96.5 & 98.1 & 97.4 & 98.4 & 97.8 & 98.6 & 96.5 & 97.4 & 97.5 & 98.3 \\
0.20 & 97.7 & 34.8 & \textbf{5.5} & \textbf{7.2} & \cellcolor[HTML]{C0C0C0}\textbf{12.6} & \textbf{16.5} & 24.8 & 51.1 & 44.4 & 75.5 & 74.1 & 77.4 & 84.1 & 89.4 & 87.6 & 84.8 & 90.6 & 85.4 & 88.2 & 91.6 & 93.3 \\
0.25 & 98.2 & 37.4 & \textbf{3.3} & \textbf{3.8} & \textbf{4.6} & \cellcolor[HTML]{C0C0C0}\textbf{5.6} & \textbf{8.7} & \textbf{13.7} & \textbf{13.6} & 28.4 & 32.5 & 38.0 & 48.2 & 58.2 & 62.5 & 58.0 & 71.4 & 72.0 & 75.8 & 82.1 & 85.9 \\
0.30 & 98.5 & 39.9 & \textbf{1.9} & \textbf{2.3} & \textbf{2.0} & \textbf{3.0} & \cellcolor[HTML]{C0C0C0}\textbf{3.5} & \textbf{3.9} & \textbf{3.7} & \textbf{10.1} & \textbf{11.0} & \textbf{14.9} & \textbf{21.4} & 28.2 & 37.7 & 35.5 & 48.6 & 55.9 & 60.4 & 69.5 & 75.9 \\
0.35 & 98.7 & 42.2 & \textbf{1.2} & \textbf{1.3} & \textbf{1.1} & \textbf{1.5} & \textbf{1.8} & \cellcolor[HTML]{C0C0C0}\textbf{1.1} & \textbf{0.9} & \textbf{3.8} & \textbf{4.0} & \textbf{5.7} & \textbf{9.3} & \textbf{12.9} & \textbf{19.8} & \textbf{19.7} & 26.8 & 38.1 & 43.6 & 53.4 & 62.3 \\
0.40 & 98.9 & 44.9 & \textbf{0.8} & \textbf{0.8} & \textbf{0.7} & \textbf{0.9} & \textbf{1.2} & \textbf{0.7} & \cellcolor[HTML]{C0C0C0}\textbf{0.4} & \textbf{1.6} & \textbf{1.2} & \textbf{2.2} & \textbf{3.6} & \textbf{5.2} & \textbf{9.4} & \textbf{9.1} & \textbf{12.8} & \textbf{21.4} & 26.7 & 35.9 & 46.8 \\
0.45 & 99.2 & 46.6 & \textbf{0.7} & \textbf{0.7} & \textbf{0.5} & \textbf{0.5} & \textbf{0.7} & \textbf{0.4} & \textbf{0.2} & \cellcolor[HTML]{C0C0C0}\textbf{0.8} & \textbf{0.6} & \textbf{0.8} & \textbf{1.6} & \textbf{2.2} & \textbf{4.0} & \textbf{4.0} & \textbf{6.0} & \textbf{10.5} & \textbf{14.2} & \textbf{19.4} & 30.7 \\
0.50 & 99.3 & 48.1 & \textbf{0.4} & \textbf{0.6} & \textbf{0.3} & \textbf{0.2} & \textbf{0.5} & \textbf{0.2} & \textbf{0.1} & \textbf{0.5} & \cellcolor[HTML]{C0C0C0}\textbf{0.3} & \textbf{0.4} & \textbf{0.7} & \textbf{1.2} & \textbf{1.8} & \textbf{1.8} & \textbf{2.7} & \textbf{4.9} & \textbf{6.9} & \textbf{9.1} & \textbf{17.9} \\
0.55 & 99.4 & 49.6 & \textbf{0.3} & \textbf{0.3} & \textbf{0.2} & \textbf{0.3} & \textbf{0.4} & \textbf{0.2} & \textbf{0.1} & \textbf{0.3} & \textbf{0.2} & \cellcolor[HTML]{C0C0C0}\textbf{0.2} & \textbf{0.4} & \textbf{0.7} & \textbf{0.9} & \textbf{0.9} & \textbf{1.5} & \textbf{2.1} & \textbf{3.4} & \textbf{4.0} & \textbf{10.0} \\
0.60 & 99.5 & 50.5 & \textbf{0.2} & \textbf{0.3} & \textbf{0.2} & \textbf{0.1} & \textbf{0.3} & \textbf{0.1} & \textbf{0.1} & \textbf{0.2} & \textbf{0.2} & \textbf{0.1} & \cellcolor[HTML]{C0C0C0}\textbf{0.3} & \textbf{0.5} & \textbf{0.5} & \textbf{0.5} & \textbf{0.7} & \textbf{1.0} & \textbf{1.6} & \textbf{1.8} & \textbf{5.7} \\
0.65 & 99.6 & 51.9 & \textbf{0.2} & \textbf{0.2} & \textbf{0.1} & \textbf{0.0} & \textbf{0.1} & \textbf{0.1} & \textbf{0.1} & \textbf{0.1} & \textbf{0.1} & \textbf{0.1} & \textbf{0.2} & \cellcolor[HTML]{C0C0C0}\textbf{0.3} & \textbf{0.3} & \textbf{0.3} & \textbf{0.5} & \textbf{0.5} & \textbf{1.0} & \textbf{0.9} & \textbf{3.1} \\
0.70 & 99.6 & 52.4 & \textbf{0.2} & \textbf{0.2} & \textbf{0.1} & \textbf{0.1} & \textbf{0.1} & \textbf{0.1} & \textbf{0.1} & \textbf{0.1} & \textbf{0.1} & \textbf{0.1} & \textbf{0.1} & \textbf{0.2} & \cellcolor[HTML]{C0C0C0}\textbf{0.3} & \textbf{0.2} & \textbf{0.4} & \textbf{0.3} & \textbf{0.6} & \textbf{0.4} & \textbf{2.1} \\
0.75 & 99.6 & 52.8 & \textbf{0.1} & \textbf{0.1} & \textbf{0.0} & \textbf{0.0} & \textbf{0.1} & \textbf{0.1} & \textbf{0.1} & \textbf{0.1} & \textbf{0.1} & \textbf{0.1} & \textbf{0.1} & \textbf{0.1} & \textbf{0.2} & \cellcolor[HTML]{C0C0C0}\textbf{0.1} & \textbf{0.2} & \textbf{0.3} & \textbf{0.3} & \textbf{0.3} & \textbf{1.4} \\
0.80 & 99.7 & 52.5 & \textbf{0.1} & \textbf{0.1} & \textbf{0.0} & \textbf{0.0} & \textbf{0.1} & \textbf{0.1} & \textbf{0.1} & \textbf{0.2} & \textbf{0.2} & \textbf{0.2} & \textbf{0.1} & \textbf{0.2} & \textbf{0.2} & \textbf{0.2} & \cellcolor[HTML]{C0C0C0}\textbf{0.1} & \textbf{0.2} & \textbf{0.2} & \textbf{0.3} & \textbf{1.4} \\
0.85 & 99.7 & 52.5 & \textbf{0.1} & \textbf{0.1} & \textbf{0.1} & \textbf{0.1} & \textbf{0.2} & \textbf{0.2} & \textbf{0.2} & \textbf{0.3} & \textbf{0.3} & \textbf{0.3} & \textbf{0.3} & \textbf{0.3} & \textbf{0.3} & \textbf{0.2} & \textbf{0.1} & \cellcolor[HTML]{C0C0C0}\textbf{0.1} & \textbf{0.2} & \textbf{0.2} & \textbf{0.7} \\
0.90 & 99.7 & 51.9 & \textbf{0.1} & \textbf{0.1} & \textbf{0.1} & \textbf{0.1} & \textbf{0.2} & \textbf{0.2} & \textbf{0.2} & \textbf{0.4} & \textbf{0.5} & \textbf{0.5} & \textbf{0.5} & \textbf{0.5} & \textbf{0.3} & \textbf{0.3} & \textbf{0.2} & \textbf{0.1} & \cellcolor[HTML]{C0C0C0}\textbf{0.2} & \textbf{0.2} & \textbf{0.3} \\
0.95 & 99.7 & 52.1 & \textbf{0.1} & \textbf{0.2} & \textbf{0.2} & \textbf{0.2} & \textbf{0.4} & \textbf{0.5} & \textbf{0.4} & \textbf{0.6} & \textbf{0.5} & \textbf{0.8} & \textbf{0.8} & \textbf{0.8} & \textbf{0.6} & \textbf{0.4} & \textbf{0.3} & \textbf{0.3} & \textbf{0.2} & \cellcolor[HTML]{C0C0C0}\textbf{0.2} & \textbf{0.3} \\
1.00 & 99.7 & 51.5 & \textbf{0.1} & \textbf{0.2} & \textbf{0.3} & \textbf{0.4} & \textbf{0.7} & \textbf{0.7} & \textbf{0.7} & \textbf{0.9} & \textbf{1.2} & \textbf{1.1} & \textbf{1.1} & \textbf{1.2} & \textbf{1.0} & \textbf{0.7} & \textbf{0.3} & \textbf{0.3} & \textbf{0.3} & \textbf{0.2} & \cellcolor[HTML]{C0C0C0}\textbf{0.2} \\
\bottomrule
\end{tabular}}
\end{table*}


\begin{table*}[]
\caption{$P_{e}$ (in\%) with a non-uniform quantization on $\beta$.
(Wave: the case of conventional embedding and detection.
Bold: the case when the steganalyst has a worse performance compared to the case of conventional embedding and detection.
Shade: the case when the steganalyst uses a matched parameter for detection, i.e., $\beta_D=\beta_A$.
Dash underline: the case when the steganalyst has a better performance with a mismatched parameter than with a matched parameter.
Underline: worst case solution for steganalyst.)}
\centering %
\label{tab:betanu_pe}       
\setlength{\tabcolsep}{0.33mm}{
\begin{tabular}{cccccccccccccccccccccc}
\toprule
$\beta_{A} \verb|\| \beta_{D}$ & 0.00 & 0.02 & 0.04 & 0.05 & 0.06 & 0.07 & 0.08 & 0.10 & 0.12 & 0.14 & 0.16 & 0.18 & 0.20 & 0.30 & 0.40 & 0.50 & 0.60 & 0.70 & 0.80 & 0.90 & 1.00 \\
\noalign{\smallskip}\hline\noalign{\smallskip}
0.00 & \cellcolor[HTML]{C0C0C0}{\uwave{20.9}} & \textbf{21.5} & \textbf{23.0} & \textbf{26.4} & \uline{\textbf{32.1}} & \textbf{37.5} & \textbf{43.2} & \textbf{50.5} & \textbf{50.9} & \textbf{50.6} & \textbf{50.9} & \textbf{50.3} & \textbf{50.1} & \textbf{50.1} & \textbf{50.0} & \textbf{50.0} & \textbf{50.0} & \textbf{50.0} & \textbf{50.0} & \textbf{50.0} & \textbf{50.0} \\
0.02 & \textbf{27.1} & \cellcolor[HTML]{C0C0C0}\textbf{23.6} & \textbf{24.3} & \textbf{26.6} & \textbf{30.2} & \textbf{33.9} & \textbf{39.1} & \textbf{49.2} & \textbf{49.7} & \textbf{50.0} & \textbf{50.4} & \textbf{50.0} & \textbf{49.9} & \textbf{50.0} & \textbf{50.0} & \textbf{50.0} & \textbf{50.0} & \textbf{50.0} & \textbf{50.0} & \textbf{50.0} & \textbf{50.0} \\
0.04 & \textbf{34.9} & \textbf{27.1} & \cellcolor[HTML]{C0C0C0}\textbf{25.7} & \textbf{26.7} & \textbf{28.3} & \textbf{30.0} & \textbf{33.1} & \textbf{45.7} & \textbf{46.6} & \textbf{48.8} & \textbf{49.1} & \textbf{49.4} & \textbf{49.7} & \textbf{49.8} & \textbf{49.9} & \textbf{50.0} & \textbf{50.0} & \textbf{50.0} & \textbf{50.0} & \textbf{50.0} & \textbf{50.0} \\
0.05 & \textbf{40.0} & \textbf{28.4} & {\dashuline{\textbf{26.3}}} & \cellcolor[HTML]{C0C0C0}\textbf{26.7} & \textbf{27.1} & \textbf{28.2} & \textbf{29.7} & \textbf{42.7} & \textbf{43.7} & \textbf{47.4} & \textbf{47.5} & \textbf{48.8} & \textbf{49.5} & \textbf{49.7} & \textbf{49.8} & \textbf{49.9} & \textbf{50.0} & \textbf{50.0} & \textbf{50.0} & \textbf{50.0} & \textbf{50.0} \\
0.06 & \textbf{42.7} & \textbf{30.3} & \textbf{27.0} & \textbf{26.7} & \cellcolor[HTML]{C0C0C0}\textbf{26.1} & \textbf{26.6} & \textbf{26.2} & \textbf{38.9} & \textbf{40.4} & \textbf{45.5} & \textbf{45.7} & \textbf{48.3} & \textbf{49.1} & \textbf{49.6} & \textbf{49.7} & \textbf{49.9} & \textbf{50.0} & \textbf{50.0} & \textbf{50.0} & \textbf{50.0} & \textbf{50.0} \\
0.07 & \textbf{46.5} & \textbf{32.3} & \textbf{28.0} & \textbf{26.8} & \textbf{25.0} & \cellcolor[HTML]{C0C0C0}\textbf{24.9} & \textbf{23.1} & \textbf{34.6} & \textbf{35.4} & \textbf{43.1} & \textbf{42.6} & \textbf{47.3} & \textbf{48.6} & \textbf{49.2} & \textbf{49.6} & \textbf{49.8} & \textbf{49.9} & \textbf{50.0} & \textbf{50.0} & \textbf{50.0} & \textbf{50.0} \\
0.08 & \textbf{50.1} & \textbf{34.4} & \textbf{28.8} & \textbf{27.2} & \textbf{24.8} & \textbf{23.9} & \cellcolor[HTML]{C0C0C0}20.3 & \textbf{30.3} & \textbf{31.0} & \textbf{39.4} & \textbf{38.5} & \textbf{45.8} & \textbf{48.1} & \textbf{48.8} & \textbf{49.4} & \textbf{49.8} & \textbf{49.9} & \textbf{50.0} & \textbf{50.0} & \textbf{49.9} & \textbf{50.0} \\
0.10 & \textbf{54.7} & \textbf{38.9} & \textbf{30.4} & \textbf{27.1} & \textbf{23.0} & \textbf{21.7} & {\dashuline{14.9}} & \cellcolor[HTML]{C0C0C0}20.1 & \textbf{21.3} & \textbf{29.0} & \textbf{26.8} & \textbf{39.7} & \textbf{45.0} & \textbf{47.1} & \textbf{48.7} & \textbf{49.6} & \textbf{49.9} & \textbf{49.9} & \textbf{49.9} & \textbf{49.9} & \textbf{49.9} \\
0.12 & \textbf{56.3} & \textbf{42.4} & \textbf{32.3} & \textbf{27.3} & \textbf{22.1} & 19.9 & {\dashuline{12.0}} & {\dashuline{13.2}} & \cellcolor[HTML]{C0C0C0}15.0 & 18.4 & 17.5 & \textbf{29.5} & \textbf{40.0} & \textbf{43.8} & \textbf{47.6} & \textbf{49.2} & \textbf{49.7} & \textbf{49.7} & \textbf{49.8} & \textbf{49.6} & \textbf{49.8} \\
0.14 & \textbf{56.8} & \textbf{46.6} & \textbf{34.2} & \textbf{27.8} & \textbf{21.2} & 18.9 & {\dashuline{9.5}} & {\dashuline{9.2}} & {\dashuline{10.9}} & \cellcolor[HTML]{C0C0C0}11.6 & 12.7 & 18.4 & \textbf{30.5} & \textbf{38.2} & \textbf{45.4} & \textbf{48.4} & \textbf{49.4} & \textbf{49.6} & \textbf{49.6} & \textbf{49.1} & \textbf{49.4} \\
0.16 & \textbf{57.2} & \textbf{49.7} & \textbf{36.1} & \textbf{28.4} & 20.5 & 17.9 & {\dashuline{8.2}} & {\dashuline{7.0}} & {\dashuline{8.6}} & {\dashuline{7.6}} & \cellcolor[HTML]{C0C0C0}9.5 & 11.3 & 19.3 & \textbf{30.0} & \textbf{40.4} & \textbf{46.6} & \textbf{48.6} & \textbf{48.9} & \textbf{49.1} & \textbf{48.0} & \textbf{48.8} \\
0.18 & \textbf{57.4} & \textbf{52.5} & \textbf{38.0} & \textbf{28.7} & 19.7 & 17.1 & {\dashuline{7.0}} & {\dashuline{5.6}} & {\dashuline{7.0}} & {\dashuline{5.7}} & 7.7 & \cellcolor[HTML]{C0C0C0}7.2 & 11.3 & 20.3 & \textbf{32.6} & \textbf{43.4} & \textbf{46.9} & \textbf{47.6} & \textbf{47.8} & \textbf{46.5} & \textbf{47.9} \\
0.20 & \textbf{57.7} & \textbf{54.5} & \textbf{40.0} & \textbf{28.8} & 19.2 & 16.5 & {\dashuline{6.6}} & {\dashuline{4.8}} & {\dashuline{6.2}} & {\dashuline{4.6}} & {\dashuline{6.6}} & {\dashuline{4.7}} & \cellcolor[HTML]{C0C0C0}6.9 & 12.7 & \textbf{22.3} & \textbf{37.1} & \textbf{42.1} & \textbf{43.9} & \textbf{45.3} & \textbf{44.2} & \textbf{46.7} \\
0.30 & \textbf{58.1} & \textbf{56.9} & \textbf{47.1} & \textbf{31.4} & 17.8 & 14.4 & 4.9 & 3.0 & 4.3 & 2.9 & 3.8 & 2.0 & {\dashuline{1.6}} & \cellcolor[HTML]{C0C0C0}2.0 & 2.0 & 5.5 & 10.7 & 18.9 & \textbf{24.4} & \textbf{30.3} & \textbf{38.0} \\
0.40 & \textbf{58.3} & \textbf{57.7} & \textbf{51.5} & \textbf{33.9} & 16.7 & 13.2 & 4.3 & 2.5 & 3.7 & 2.5 & 2.8 & 1.8 & 0.9 & 0.9 & \cellcolor[HTML]{C0C0C0}0.3 & 0.7 & 1.8 & 4.8 & 6.5 & 13.4 & \textbf{23.5} \\
0.50 & \textbf{58.5} & \textbf{58.2} & \textbf{53.8} & \textbf{35.5} & 16.4 & 12.6 & 4.0 & 2.3 & 3.2 & 2.2 & 2.4 & 1.7 & 0.7 & 0.5 & 0.2 & \cellcolor[HTML]{C0C0C0}0.2 & 0.4 & 1.0 & 1.4 & 3.5 & 9.0 \\
0.60 & \textbf{58.6} & \textbf{58.5} & \textbf{54.9} & \textbf{36.7} & 17.2 & 12.2 & 3.8 & 2.2 & 3.1 & 2.2 & 2.1 & 1.7 & 0.7 & 0.4 & 0.2 & {\dashuline{0.1}} & \cellcolor[HTML]{C0C0C0}0.2 & 0.3 & 0.4 & 0.8 & 2.9 \\
0.70 & \textbf{58.6} & \textbf{58.9} & \textbf{55.2} & \textbf{37.6} & 17.5 & 12.0 & 3.8 & 2.1 & 3.0 & 2.1 & 1.9 & 1.6 & 0.6 & 0.3 & 0.2 & {\dashuline{0.1}} & {\dashuline{0.1}} & \cellcolor[HTML]{C0C0C0}0.2 & 0.2 & 0.4 & 1.1 \\
0.80 & \textbf{58.7} & \textbf{59.0} & \textbf{55.1} & \textbf{37.7} & 18.2 & 11.8 & 3.8 & 2.1 & 3.0 & 2.1 & 1.8 & 1.6 & 0.6 & 0.3 & 0.2 & 0.2 & 0.1 & 0.1 & \cellcolor[HTML]{C0C0C0}0.1 & 0.1 & 0.6 \\
0.90 & \textbf{58.7} & \textbf{59.1} & \textbf{54.2} & \textbf{37.4} & 18.6 & 11.9 & 3.8 & 2.1 & 3.0 & 2.1 & 1.7 & 1.7 & 0.6 & 0.4 & 0.2 & 0.3 & 0.3 & 0.2 & 0.1 & \cellcolor[HTML]{C0C0C0}0.1 & 0.3 \\
1.00 & \textbf{58.7} & \textbf{58.9} & \textbf{53.0} & \textbf{37.2} & 18.3 & 11.9 & 3.9 & 2.1 & 3.0 & 2.2 & 1.7 & 1.7 & 0.7 & 0.7 & 0.5 & 0.6 & 0.6 & 0.6 & 0.2 & 0.2 & \cellcolor[HTML]{C0C0C0}0.2 \\
\bottomrule
\end{tabular}}
\end{table*}
%
%
\begin{table*}
\vspace{0.2cm}
\caption{$P_{md}$ with a non-uniform quantization on $\beta$.
(Wave: the case of conventional embedding and detection.
Shade: the case when the steganalyst uses a matched parameter for detection, i.e., $\beta_D=\beta_A$.
Bold: the case when the steganographer has a better performance compared to the case of conventional embedding and detection.
Double underline: worst case solution for steganographer.)}
\centering %
\label{tab:betanu_pmd}       
\setlength{\tabcolsep}{0.35mm}{
\begin{tabular}{cccccccccccccccccccccc}
\toprule
$\beta_{A} \verb|\| \beta_{D}$ & 0.00 & 0.02 & 0.04 & 0.05 & 0.06 & 0.07 & 0.08 & 0.10 & 0.12 & 0.14 & 0.16 & 0.18 & 0.20 & 0.30 & 0.40 & 0.50 & 0.60 & 0.70 & 0.80 & 0.90 & 1.00 \\
\noalign{\smallskip}\hline\noalign{\smallskip}
0.00 & \cellcolor[HTML]{C0C0C0}\uwave{24.0} & \textbf{21.3} & 26.3 & 29.8 & 44.3 & 55.8 & 79.1 & 96.8 & 96.2 & 97.1 & 98.8 & 97.4 & 99.1 & 99.5 & 99.8 & 100.0 & 99.9 & 99.9 & 99.9 & 99.9 & 99.9 \\
0.02 & 36.4 & \cellcolor[HTML]{C0C0C0}25.5 & 28.9 & 30.3 & 40.4 & 48.5 & 70.8 & 94.3 & 93.9 & 95.8 & 97.7 & 96.7 & 98.7 & 99.4 & 99.7 & 99.8 & 99.9 & 99.9 & 99.9 & 99.9 & 99.9 \\
0.04 & 52.2 & 32.5 & \cellcolor[HTML]{C0C0C0}31.7 & \uuline{30.6} & 36.7 & 40.7 & 58.8 & 87.2 & 87.6 & 93.4 & 95.2 & 95.6 & 98.2 & 99.0 & 99.5 & 99.8 & 99.9 & 99.9 & 99.9 & 99.9 & 99.9 \\
0.05 & 60.3 & 35.1 & 32.8 & \cellcolor[HTML]{C0C0C0}\uuline{30.6} & 34.2 & 37.1 & 52.0 & 81.3 & 81.7 & 90.7 & 92.1 & 94.3 & 97.8 & 98.8 & 99.3 & 99.8 & 99.8 & 99.9 & 99.9 & 99.8 & 99.9 \\
0.06 & 67.7 & 38.9 & 34.3 & 30.5 & \cellcolor[HTML]{C0C0C0}32.1 & 34.0 & 45.2 & 73.7 & 75.2 & 86.8 & 88.5 & 93.3 & 97.0 & 98.6 & 99.2 & 99.7 & 99.9 & 99.8 & 99.9 & 99.8 & 99.8 \\
0.07 & 75.4 & 42.9 & 36.3 & 30.6 & 30.0 & \cellcolor[HTML]{C0C0C0}30.6 & 38.8 & 65.1 & 65.1 & 82.0 & 82.2 & 91.4 & 96.0 & 97.9 & 98.9 & 99.5 & 99.8 & 99.8 & 99.9 & 99.8 & 99.8 \\
0.08 & 82.4 & 47.1 & 37.9 & 31.6 & 29.5 & 28.6 & \cellcolor[HTML]{C0C0C0}33.2 & 56.5 & 56.4 & 74.7 & 74.0 & 88.4 & 95.0 & 97.0 & 98.5 & 99.5 & 99.8 & 99.8 & 99.8 & 99.7 & 99.8 \\
0.10 & 91.7 & 56.0 & 41.0 & 31.2 & 26.0 & 24.2 & \textbf{22.6} & \cellcolor[HTML]{C0C0C0}36.1 & 36.9 & 53.9 & 50.7 & 76.1 & 88.8 & 93.6 & 97.1 & 99.2 & 99.6 & 99.7 & 99.7 & 99.6 & 99.7 \\
0.12 & 94.9 & 63.1 & 45.0 & 31.6 & 24.1 & \textbf{20.6} & \textbf{16.8} & \textbf{22.4} & \cellcolor[HTML]{C0C0C0}24.4 & 32.7 & 32.0 & 55.7 & 78.9 & 86.9 & 94.9 & 98.2 & 99.4 & 99.3 & 99.5 & 99.1 & 99.5 \\
0.14 & 96.0 & 71.4 & 48.7 & 32.7 & \textbf{22.5} & \textbf{18.6} & \textbf{11.8} & \textbf{14.3} & \textbf{16.2} & \cellcolor[HTML]{C0C0C0}\textbf{19.1} & \textbf{22.5} & 33.6 & 60.0 & 75.8 & 90.5 & 96.7 & 98.6 & 99.0 & 99.1 & 98.1 & 98.7 \\
0.16 & 96.7 & 77.7 & 52.4 & 33.9 & \textbf{20.9} & \textbf{16.6} & \textbf{9.1} & \textbf{9.8} & \textbf{11.6} & \textbf{11.1} & \cellcolor[HTML]{C0C0C0}\textbf{15.9} & \textbf{19.4} & 37.4 & 59.4 & 80.4 & 93.2 & 97.2 & 97.8 & 98.1 & 95.8 & 97.4 \\
0.18 & 97.1 & 83.2 & 56.2 & 34.5 & \textbf{19.5} & \textbf{14.9} & \textbf{6.8} & \textbf{7.2} & \textbf{8.4} & \textbf{7.2} & \textbf{12.3} & \cellcolor[HTML]{C0C0C0}\textbf{11.1} & \textbf{21.4} & 40.0 & 64.9 & 86.8 & 93.6 & 95.0 & 95.5 & 92.8 & 95.8 \\
0.20 & 97.7 & 87.2 & 60.3 & 34.8 & \textbf{18.4} & \textbf{13.7} & \textbf{5.8} & \textbf{5.5} & \textbf{6.9} & \textbf{5.1} & \textbf{10.2} & \textbf{6.2} & \cellcolor[HTML]{C0C0C0}\textbf{12.6} & \multicolumn{1}{c}{24.8} & 44.4 & 74.1 & 84.1 & 87.6 & 90.6 & 88.2 & 93.3 \\
0.30 & 98.5 & 92.1 & 74.5 & 39.9 & \textbf{15.7} & \textbf{9.5} & \textbf{2.5} & \textbf{1.9} & \textbf{2.9} & \textbf{1.7} & \textbf{4.5} & \textbf{0.8} & \textbf{2.0} & \multicolumn{1}{c}{\cellcolor[HTML]{C0C0C0}\textbf{3.5}} & \textbf{3.7} & \textbf{11.0} & \textbf{21.4} & 37.7 & 48.6 & 60.4 & 75.9 \\
0.40 & 98.9 & 93.7 & 83.2 & 44.9 & \textbf{13.3} & \textbf{7.2} & \textbf{1.3} & \textbf{0.8} & \textbf{1.7} & \textbf{0.8} & \textbf{2.7} & \textbf{0.3} & \textbf{0.7} & \multicolumn{1}{c}{\textbf{1.2}} & \cellcolor[HTML]{C0C0C0}\textbf{0.4} & \textbf{1.2} & \textbf{3.6} & \textbf{9.4} & \textbf{12.8} & 26.7 & 46.8 \\
0.50 & 99.3 & 94.6 & 87.8 & 48.1 & \textbf{12.8} & \textbf{5.9} & \textbf{0.7} & \textbf{0.4} & \textbf{0.9} & \textbf{0.4} & \textbf{1.8} & \textbf{0.2} & \textbf{0.3} & \multicolumn{1}{c}{\textbf{0.5}} & \textbf{0.1} & \cellcolor[HTML]{C0C0C0}\textbf{0.3} & \textbf{0.7} & \textbf{1.8} & \textbf{2.7} & \textbf{6.9} & \textbf{17.9} \\
0.60 & 99.5 & 95.3 & 90.0 & 50.5 & \textbf{14.4} & \textbf{5.2} & \textbf{0.4} & \textbf{0.2} & \textbf{0.6} & \textbf{0.3} & \textbf{1.1} & \textbf{0.1} & \textbf{0.2} & \multicolumn{1}{c}{\textbf{0.3}} & \textbf{0.1} & \textbf{0.2} & \cellcolor[HTML]{C0C0C0}\textbf{0.3} & \textbf{0.5} & \textbf{0.7} & \textbf{1.6} & \textbf{5.7} \\
0.70 & 99.6 & 96.0 & 90.7 & 52.4 & \textbf{15.0} & \textbf{4.7} & \textbf{0.3} & \textbf{0.2} & \textbf{0.4} & \textbf{0.1} & \textbf{0.8} & \textbf{0.1} & \textbf{0.1} & \multicolumn{1}{c}{\textbf{0.1}} & \textbf{0.1} & \textbf{0.1} & \textbf{0.1} & \cellcolor[HTML]{C0C0C0}\textbf{0.3} & \textbf{0.4} & \textbf{0.6} & \textbf{2.1} \\
0.80 & 99.7 & 96.2 & 90.4 & 52.5 & \textbf{16.4} & \textbf{4.4} & \textbf{0.4} & \textbf{0.1} & \textbf{0.3} & \textbf{0.1} & \textbf{0.6} & \textbf{0.1} & \textbf{0.0} & \multicolumn{1}{c}{\textbf{0.1}} & \textbf{0.1} & \textbf{0.2} & \textbf{0.1} & \textbf{0.2} & \cellcolor[HTML]{C0C0C0}\textbf{0.1} & \textbf{0.2} & \textbf{1.1} \\
0.90 & 99.7 & 96.5 & 88.7 & 51.9 & \textbf{17.2} & \textbf{4.6} & \textbf{0.4} & \textbf{0.1} & \textbf{0.3} & \textbf{0.1} & \textbf{0.4} & \textbf{0.1} & \textbf{0.1} & \multicolumn{1}{c}{\textbf{0.2}} & \textbf{0.2} & \textbf{0.5} & \textbf{0.5} & \textbf{0.3} & \textbf{0.2} & \cellcolor[HTML]{C0C0C0}\textbf{0.2} & \textbf{0.5} \\
1.00 & 99.7 & 96.1 & 86.3 & 51.5 & \textbf{16.6} & \textbf{4.6} & \textbf{0.5} & \textbf{0.1} & \textbf{0.3} & \textbf{0.2} & \textbf{0.3} & \textbf{0.2} & \textbf{0.3} & \multicolumn{1}{c}{\textbf{0.7}} & \textbf{0.7} & \textbf{1.2} & \textbf{1.1} & \textbf{1.0} & \textbf{0.3} & \textbf{0.3} & \cellcolor[HTML]{C0C0C0}\textbf{0.3} \\
\bottomrule
\end{tabular}}
\end{table*}

\section{Experimental Results}
\label{sec:exp}

In this section, we report the {results of our experiments and discuss their meaning.

\subsection{Payoff matrices}
\label{sec:exp:PAYOFF}



{ In Table \ref{tab:betaU_pe}, we report the payoff matrix $\mathbf P_{e}$ obtained from our experiments
when  $\beta_A$ and $\beta_D$  are quantized with a uniform quantization step size equal to 0.05. We also report the payoff matrix $\mathbf P_{md}$ in Table \ref{tab:betaU_pmd}. 
From the analysis carried out in Section \ref{sec:game} (and the result stated by Property \ref{property}), we know that only the matrix $\mathbf P_{e}$ is necessary to solve the game and derive the optimum strategies for D and A. Then, the values of $P_{md}$ corresponding to the optimum parameters  are used to compute the payoff of A at the equilibrium.}
%
%
In the tables, we highlight the results for the case $\beta_A=\beta_D$ (diagonal) with grey shades.
The performance
at $(\beta_A, \beta_D) =(0,0)$ corresponds to the performance when conventional S-UNIWARD steganography and steganalysis are {considered (this case provides the baseline performance and is highlighted with a wave line)}, in which case  $P_e=20.9\%$ and $P_{md}=24.0\%$.
%
%
%

For a fixed $\beta_A$, one would expect that the best possible performance for D is achieved when $\beta_D=\beta_A$.
Actually, this is not always the case.
%
However, in all the cases where, for a given $\beta_A$, the steganalyzer $\phi_{\mathcal{C}_{trn},\mathcal{Z}^{\beta_A}_{trn}}$ is outperformed by
$\phi_{\mathcal{C}_{trn},\mathcal{Z}^{\beta_D}_{trn}}$ with a $\beta_D \neq \beta_A$
{(highlighted with dash underlines in the table),}
the difference in the corresponding value of the $P_e$  is not much.
It can also be observed that the region where $0.1 \le \beta_D <\beta_A \le 0.3$ is where the difference between the performance of the steganalyzer with matched and mismatched $\beta$ is more relevant.
By closer inspection of Table \ref{tab:betaU_pe}, we also observe that, in this region, the performance of the steganalyzer varies greatly, thus calling for a further investigation on a finer quantization of $\beta$.
{For this reason}, we also considered a different quantization for the $\beta$ values, which is finer for small $\beta$, and coarser for large $\beta$; in particular, we considered the following non-uniform set of values: $\mathcal{S}^{q}_{nu}=$ $\{$0, 0.02, 0.04, 0.05, 0.06, 0.07, 0.08, 0.10, 0.12, 0.14, 0.16, 0.18, 0.20, 0.30, 0.40, 0.50, 0.60, 0.70, 0.80, 0.90, 1.00$\}$.
The corresponding matrices $\mathbf P_{e}$ and $\mathbf P_{md}$  are shown in Table \ref{tab:betanu_pe} and \ref{tab:betanu_pmd}, respectively.

{From both} Table \ref{tab:betaU_pe} and \ref{tab:betanu_pe}, we can observe the following general behavior:
\begin{itemize}
  \item when D uses a steganalyzer with a very small $\beta_D$ ($\beta_D \le 0.05$),
the attacker {would better choose} a large $\beta_A$;
  \item when D uses a steganalyzer with a large $\beta_D$,
for the attacker it is better to choose a small $\beta_A$;
\item for matched values of $\beta_A$ and $\beta_D$ (corresponding to the diagonal of the matrix), the attacker can outperform the baseline only when such values are relatively small.
\end{itemize}
A similar behavior also holds for the $\mathbf P_{md}$ matrix in Table  \ref{tab:betaU_pmd} and \ref{tab:betanu_pmd}.

\begin{table*}[]
\vspace{0.2cm}
\caption{The mixed strategy equilibrium for uniform quantization on $\beta$ with a step of 0.05. The corresponding payoff matrices $\mathbf P_{e}$ and $\mathbf P_{md}$ were shown in Table \ref{tab:betaU_pe} and \ref{tab:betaU_pmd}, respectively.}
\centering %
\label{tab:betaU_mix_nzero}       
\setlength{\tabcolsep}{0.42mm}{
\begin{tabular}{@{}lccccccccccccccccccccc@{}}
\toprule
$\beta$ & 0.00 & 0.05 & 0.10 & 0.15 & 0.20 & 0.25 & 0.30 & 0.35 & 0.40 & 0.45 & 0.50 & 0.55 & 0.60 & 0.65 & 0.70 & 0.75 & 0.80 & 0.85 & 0.90 & 0.95 & 1.00 \\ \midrule
$p^{*}_A(\beta_{A})$ & 0.476 & 0.140    & 0    & 0     & 0     & 0    & 0    & 0    & 0    & 0    & 0    & 0    & 0    & 0    & 0    & 0.384    & 0    & 0    & 0    & 0 & 0    \\
$p^{*}_D(\beta_{D})$ & 0     & 0.812    & 0.040    & 0 & 0 & 0    & 0    & 0    & 0    & 0    & 0.148    & 0    & 0    & 0    & 0    & 0    & 0    & 0    & 0    & 0     & 0    \\
\noalign{\smallskip}\hline\noalign{\smallskip}
$P_{md}^*$      &       &      &      &       &       &      &      &      &      &      & 42.9\% &      &      &      &      &      &      &      &      &       &      \\
$P_e^*$      &       &      &      &       &       &      &      &      &      &      & 30.8\% &      &      &      &      &      &      &      &      &       &      \\ \bottomrule
\end{tabular}}
\end{table*}
\begin{table*}[]
\vspace{0.2cm}
\caption{The mixed strategy equilibrium for non-uniform quantization on $\beta$. The corresponding payoff matrices $\mathbf P_{e}$ and $\mathbf P_{md}$  were shown in Table \ref{tab:betanu_pe} and \ref{tab:betanu_pmd}, respectively.}
\centering %
\label{tab:betanu_nzero}       
\setlength{\tabcolsep}{0.42mm}{
\begin{tabular}{@{}cccccccccccccccccccccc@{}}
\toprule
$\beta$ & 0.00  & 0.02 & 0.04 & 0.05  & 0.06  & 0.07 & 0.08 & 0.10 & 0.12 & 0.14 & 0.16 & 0.18 & 0.20 & 0.30 & 0.40 & 0.50 & 0.60 & 0.70 & 0.80 & 0.90  & 1.00 \\ \midrule
$p^{*}_A(\beta_{A})$ & 0.767 & 0    & 0    & 0     & 0     & 0    & 0    & 0    & 0    & 0    & 0    & 0    & 0    & 0    & 0    & 0    & 0    & 0    & 0    & 0.233 & 0    \\
$p^{*}_D(\beta_{D})$ & 0     & 0    & 0    & 0.551 & 0.449 & 0    & 0    & 0    & 0    & 0    & 0    & 0    & 0    & 0    & 0    & 0    & 0    & 0    & 0    & 0     & 0    \\
\noalign{\smallskip}\hline\noalign{\smallskip}
$P_{md}^*$      &       &      &      &       &       &      &      &      &      &      & 36.3\% &      &      &      &      &      &      &      &      &       &      \\
$P_e^*$      &       &      &      &       &       &      &      &      &      &      & 29.0\% &      &      &      &      &      &      &      &      &       &      \\ \bottomrule
\end{tabular}}
\end{table*}

\subsection{Equilibrium of the game} 
\label{sec:exp:NASH}

In Table \ref{tab:betaU_mix_nzero} we report the mixed strategy Nash equilibrium $(\bm{p}_A^*, \bm{p}_D^*)$ for the $ASED$ game when
$\beta$ is uniformly quantized (there are no equilibria in pure strategies).
At the equilibrium, the optimal payoff for the defender is $P^*_{e}=30.8\%$, while for the attacker the optimal payoff is $P^*_{md}=42.9\%$.
Compared to conventional embedding where $P_{e}(0, 0)=20.9\%$, the error probability is increased by almost 10\%.
This percentage characterizes the loss in the error probability with respect to the case of conventional
non-adversarial steganography, due to adversarial embedding.
With regard to the equilibrium strategies, interestingly, it can be observed that the optimum behavior of the steganalyst, and especially the steganographer, corresponds to alternate playing a small and a medium/large $\beta$.
%
%
In particular, among the possible {values of $\beta$},
the best for the attacker is  to perform conventional steganography ($\beta_A = 0$) and adversarial steganography with $\beta_A = 0.05$, with a probability of 0.476 and 0.140 respectively, and to consider a stronger adversarial embedding with $\beta_A = 0.75$
with a probability of 0.384.
%
On the other {hand},
for the defender, the best is to consider
the aware steganalyzers  $\phi_{\mathcal{C}_{trn},\mathcal{Z}^{0.05}_{trn}}$, $\phi_{\mathcal{C}_{trn},\mathcal{Z}^{0.1}_{trn}}$, and $\phi_{\mathcal{C}_{trn},\mathcal{Z}^{0.5}_{trn}}$
with a probability of 0.812, 0.040, and 0.148, respectively.


In Table \ref{tab:betanu_nzero}, we show the equilibrium when $\beta$ is quantized non-uniformly.
The optimal payoff for the defender is $P^*_{e}=29.0\%$,
while for the attacker is $P^*_{md}=36.3\%$.
Compared to the case of uniform quantization,
using finer quantization in the region where the payoffs vary rapidly
gives  slightly more advantage to the defender. However, the behavior of the optimum mixed strategy at the equilibrium is similar to {the one observed} before: in particular, for the attacker, the best is to use a conventional steganographic scheme  with an increased probability of 0.767,  and a strong  adversarial embedding (with $\beta_A= 0.9$) for the remaining {instances}.
%
For the defender,
the probabilities are now concentrated on $\beta = [0.05, 0.06]$; in particular, D should use
$\phi_{\mathcal{C}_{trn},\mathcal{Z}^{0.05}_{trn}}$ and $\phi_{\mathcal{C}_{trn},\mathcal{Z}^{0.06}_{trn}}$,
with probability of 0.551 and 0.449, respectively.

In both cases, and especially in the case of non-uniform quantization, the shape of the equilibrium strategy of A  confirms
the necessity for the steganographer to find a good \textit{trade-off}
between hiding the adversarial embedding on one {hand} ($\beta= 0$), and trying to force the classifier towards a wrong decision (adversarial stego detected as a cover) by {using a large $\beta$}, on the other.

\subsection{Worst case solution}
\label{sec:exp:WORST}


If the steganographer and the steganalyst decide to adopt a worst case approach, as described in Section \ref{sec:::worstcase},
we obtain the following results:
\begin{itemize}
  \item for the case of uniform quantization,
the worst case strategy for A is
$\hat{\beta}_A=0.10$ yielding a worst case payoff $\hat{P}_{md}=31.2\%$ (achieved when $\beta_D=0.05$), which is highlighted {with double-underline} in Table \ref{tab:betaU_pmd},
while that for D is $\hat{\beta}_D =  0.05$,
for which the worst case payoff is $\hat{P}_{md}=37.8\%$ (achieved when $\beta_A=0.75$ and
{highlighted} {with underline} in Table \ref{tab:betaU_pe});
  \item for the case of non-uniform quantization,
the  worst case strategy for A is the profile
$\hat{\beta}_A=0.04$, or $0.05$ (the worst case payoff is $\hat{P}_e=30.6\%$),
{(highlighted} {with double-underline in Table \ref{tab:betanu_pe})},
while that for D is $\hat{\beta}_D=0.06$,
with a worst case payoff $\hat{P}_e=32.1\%$ (highlighted {with underline} in Table \ref{tab:betanu_pe}).
\end{itemize}
Therefore, in both cases, we have $\hat{P}_e > P_e^*$ and $\hat{P}_{md} < P_{md}^*$; hence, as expected, for both $D$ and $A$ the worst case solution leads to a smaller (significantly smaller in some cases) payoff compared to the Nash equilibrium, thus confirming the benefit of {adopting} the Nash equilibrium solution.

Our results show that the quantization of $\beta$ plays an important role {to determine} the result of the game.
%
%
Notably, the results do not change significantly by using a finer quantization with respect to the one considered in the set $\mathcal{S}^{q}_{nu}$, thus indicating that the sampling in
$\mathcal{S}^{q}_{nu}$ is already dense enough.

\subsection{Performance of a steganalyzer trained over a mixture of $\beta_D$}
\label{sec:exp:MIX}
In the previous section, we have obtained experimentally the mixed strategy Nash equilibriums for the $ASED$ game, where $\beta$ can take values in a finite set of values.
{We considered a situation wherein} the steganalyzer was trained on a mixture of adversarial stego images, generated with different $\beta$.
Two compositions for the training set were considered:
\begin{enumerate}
  \item ${\mathcal{Z}^{\text{uni}}_{trn}}$: this set was consisted of adversarial images with $\beta$ distributed according to a uniform distribution.
  For example, in the case of uniform quantization of $\beta$, {21 values of $\beta$ were possible}, then each $\beta$ contributed with about 190 images (the whole training set {consisting of} $4000$ images);
  \item ${\mathcal{Z}^{\mathbfit{p}^*_D}_{trn}}$: this set was consisted of adversarial images  with $\beta$ distributed according to the probability distribution $\mathbfit{p}^*_D$ at the equilibrium of the $ASED$ game. For example, in the case of uniform quantization, the training set contained 3252, 120, and 628 adversarial images with $\beta = 0.05$, $0.10$, and $0.50$, respectively (see Table \ref{tab:betaU_mix_nzero}).
\end{enumerate}
The adversarial stego images and their cover counterparts formed the training set for the steganalyzer.
The validation sets were composed by images in the same proportion.
The same test set introduced in Section \ref{sec:methodology:SET} was used for testing.
The results of the tests are given in Table \ref{tab:uniform}
for the case of uniform and non-uniform quantization.
%
For each case,  the table reports  the value of the error probability weighted according to the probability distribution $\mathbfit{p}^*_A$ at the equilibrium, indicated by $\widetilde{P}_{e}$, achieved by the classifiers $\phi_{\mathcal{C}_{trn},\mathcal{Z}^{\text{uni}}_{trn}}$ and  $\phi_{\mathcal{C}_{trn},\mathcal{Z}^{\mathbfit{p}^*_D}_{trn}}$.

\begin{table}[]
\small
\caption{Performance of the steganalyzers trained on mixtures of adversarial stego images with different $\beta$, for the case of uniform and non-uniform quantization on $\beta$.}
\vspace{0.3cm}
\centering %
\label{tab:uniform}       
\begin{tabular}{@{}cccccc@{}}
\toprule
&  \multicolumn{2}{c}{uniform quantization} & \multicolumn{2}{c}{non-uniform  quantization}\\
&  ${\mathcal{Z}^{\text{uni}}_{trn}}$  & $\mathcal{Z}^{\mathbfit{p}^*_D}_{trn}$ &     ${\mathcal{Z}^{\text{uni}}_{trn}}$  & $\mathcal{Z}^{\mathbfit{p}^*_D}_{trn}$ &    \\
\midrule
$\widetilde{P}_{e}$     & 26.0   & \textbf{20.7}  & 30.2  & \textbf{28.8}   \\
\bottomrule
\end{tabular}
\end{table}

In the case of uniform quantization of $\beta$,
it can be observed that
the steganalyzer trained with $\{ \mathcal{C}_{trn}, {\mathcal{Z}^{\mathbfit{p}^*_D}_{trn}} \}$ leads to the best performance for D.
Moreover, we observe that, when using $\{ \mathcal{C}_{trn}, {\mathcal{Z}^{\mathbfit{p}^*_D}_{trn}} \}$ for training, the resulting $\widetilde{P}_{e} = 20.7\%$ is much lower than the value at the Nash equilibrium (which is, ${P}_{e}^{*}= 30.8\%$),

Similarly, in the case of non-uniform quantization,
it can be observed that the classifier trained on $\{ \mathcal{C}_{trn}, {\mathcal{Z}^{\mathbfit{p}^*_D}_{trn}} \}$
achieves the best detection performance.
%
When the attacker plays the equilibrium strategy, the steganalyzer trained with $\{ \mathcal{C}_{trn}, {\mathcal{Z}^{\mathbfit{p}^*_D}_{trn}} \}$ achieves better performance with respect to the Nash equilibrium ($\widetilde{P}_{e}=28.8\%$ and ${P}_{e}^{*}=29.0\%$, respectively), though the improvement in this case is not as strong as in the previous case.

Overall, the above results confirm that D can get an advantage by training the steganalyzer with a mixture of images distributed as in the distribution provided by the Nash equilibrium strategy.
Following this observation, it would be very interesting to see what happens when A changes his strategy, with respect to the equilibrium mixture, as a reaction (knowing about the mixed training adopted by D). This would naturally lead to the definition of a more general game where the set of strategies for A and D are the possible mixtures or distributions over the $\beta$ values (rather than the pure $\beta$ values, as in the definition of the $ASED$ game). Obviously, the computational burden of these tests is enormous, due to the need of training over all the possible mixtures, or, at least, a wide variety of them.
The investigation of this case is left as a future work.

\section{Conclusions}
\label{sec:conclusion}

We formulated the interplay between CNN-based steganalysis and adversarial embedding as a
two-player non-zero-sum strategic game.
In particular, the game is played on the value of the parameter $\beta$, ruling the amount of adjustable elements set by the steganographer during the attack (strength of the adversarial embedding), which the stenaganalyst tries to guess and use for adversarial training.

We have shown that the solution of the non zero-sum game can be traced back to the solution of an associated zero-sum game, for which the Nash equilibrium can be derived more easily. The experiments that we run in a practical setup of CNN-based steganalysis and adversarial embedding provide the optimum behavior for the steganalyst and the steganographer and show the performance that can be achieved by playing at the Nash equilibrium, thus characterizing the loss in the error probability with respect to the case of conventional non-adversarial steganography.
From our experiments, we can also observe that the quantization of the parameter $\beta$ plays an important role in the game solution, affecting the performance at the equilibrium.
As a further result, we verified that an improved solution for the steganalyst, i.e.,  yielding a lower error probability, can be obtained by training the classifier with the mixture of $\beta$ values provided by the steganalyst's equilibrium strategy of the game. This is an interesting result that goes beyond the game analysis considered in this paper, and thus calls for more investigation as a future work.

Other games could be defined
by considering different sets of strategies for {the two players}. For example, the selection-channel information is a kind of strategy that both parties may utilize. In this respect, it {would be} interesting to investigate the interplay between selection-channel steganalysis and content-adaptive steganography from a game-theoretic perspective.

%


\bibliographystyle{unsrt}
\bibliography{mybibfile}


\end{document}